\documentclass[11pt, a4paper]{article}

\usepackage{amsmath, amssymb}		
\usepackage{enumitem}
\setlist[description]{leftmargin=0.5cm,labelindent=0cm}
\usepackage{natbib}
\usepackage{hyperref}

\begin{document}

\title{Dimensions of
  ``Timescales'' in Neuromorphic Computing Systems}

\author{Herbert Jaeger\\University of Groningen\\h.jaeger@rug.nl \and Celestine Lawrence\\University of Groningen\\c.p.lawrence@rug.nl \and Dirk Doorakkers\\University of Groningen\\h.a.doorakkers@rug.nl \and Giacomo Indiveri\\ETH Zurich and University of Zurich\\giacomo@ini.uzh.ch}

\date{February 10, 2021}

\maketitle

\begin{abstract} This article is a public deliverable of the EU
  project \emph{Memory technologies with multi-scale time constants
    for neuromorphicarchitectures} (MeMScales,
  \href{https://memscales.eu/}{memscales.eu/}, Call ICT-06-2019
  Unconventional Nanoelectronics, project number 871371). This arXiv
  version is a verbatim copy of the deliverable report, with
  administrative information stripped. It collects a wide and varied
  assortment of phenomena, models, research themes and algorithmic
  techniques that are connected with timescale phenomena in the fields
  of computational neuroscience, mathematics, machine learning and
  computer science, with a bias toward aspects that are relevant for
  neuromorphic engineering. It turns out that this theme is very rich
  indeed and spreads out in many directions which defy a unified
  treatment. We collected several dozens of sub-themes, each of which
  has been investigated in specialized settings (in the neurosciences,
  mathematics, computer science and machine learning) and has been
  documented in its own body of literature. The more we dived into
  this diversity, the more it became clear that our first effort to
  compose a survey must remain sketchy and partial. We conclude with a
  list of insights distilled from this survey which give general
  guidelines for the design of future neuromorphic systems. \end{abstract}


\clearpage
\tableofcontents

\section{Introduction and overview}

The original topic for this deliverable, agreed more than a year ago
and specified in the project proposal, was \emph{Report on literature
  survey and analysis of STDP and RC} [= spike-time dependent
plasticity and reservoir computing, respectively]
\emph{guidance for the design of indeterminate hardware,} with the
understanding that the guiding focus of this survey would lie on
timescale aspects. When we started to carry out this survey we
perceived that, first, the restriction to STDP and RC would exclude
many technically and algorithmically important phenomena in general
analog (spiking) microchips --- so we extended our perspective to
neuromorphic computing in general. Second, we realized that the
phenomenology of ``timescales'' is very rich, and this word is applied
to quite different phenomena in different contexts. Therefore, an
important contribution of this deliverable report is to stake out the
conceptual dimensions of this scintillating word. Hence our new title,
\emph{Dimensions of ``timescales'' in neuromorphic computing systems.}

The report is structured as follows. In Section \ref{secConcepts} we
unfold the conceptual dimensions of the timescales concept, by
pointing out different uses and subconcepts of this notion in
different formal-theoretical, computational and physical contexts. In
the following three sections we compile the findings of a literature
survey, sorted into the fields of neuroscience (Section
\ref{secBrains}), mathematics and theoretical physics (Section
\ref{secMath}), and computer science / machine learning (Section
\ref{secComp}). Section \ref{secAdvice} distils a number of take-home
messages distilled from the findings in this deliverable which we hope
are helpful for informing future research in MemScales and beyond. A
final Section \ref{secGuides} gives concrete physical-timescale related
guidelines for the design of indeterminate hardware which result from
the specific givens in recent developments in STDP and RC research.

\section{Talking and thinking about timescales: dimensions of a very
  rich concept}\label{secConcepts}

In this section we give a ``travel guide'' for the landscape of
timescale phenomena, and point out terminologies that are used. We
found it not possible (at least, not at present) to develop a unified,
comprehensive conceptual framework. Therefore, we present our findings
in the form of a collection of objects and places of interest, as in a
tourist guide where showplaces are paid passing visits. 

\newcounter{tdimension}
\renewcommand{\thetdimension}{\Alph{tdimension}}

\begin{description}
   \refstepcounter{tdimension} \label{speedMemory}
\item[\Alph{tdimension} ``Speed'' and ``memory''.] There are at least
  two different, but likewise fundamental, understandings of
  ``timescales''.

  The first one is
  to speak of fast or slow timescales when a dynamical system evolves
  faster or slower, as one could for instance mathematically determine
  by changing \emph{time constants} in ordinary differential equations
  (ODEs). When the system is ``fast'', its rates of change in
  numerical dynamical variables are high --- timeseries will exhibit
  many high absolute first derivatives, have strong components in the
  high-frequency end of its Fourier spectrum, etc. We note, however,
  that there is no unique mathematical criterion to measure
  ``speed''. For instance, if an ODE-defined dynamical system (DS) is
  close or in a fixed point attractor, even very small ODE time
  constants will not translate to large numerical change rates.

  The second is to speak of long or short \emph{memory} durations. In
  this view, a DS is evolving on a slow timescale when it has long
  ``memory spans''. This means that some information that is
  \emph{encoded} in the system state at some time $t$ can be decoded
  again at (much) later times. Instead of using the word ``memory'',
  which is too closely suggestive of neural and cognitive processing,
  we find it preferable to speak of ``preservation of information
  across time''. Exploring the preservation of information across time
  has been one of the main themes in the theoretical literature on
  reservoir computing (RC) in the last 20 years.
 \refstepcounter{tdimension} \label{stateControlVar}
\item[\Alph{tdimension} State variables, control parameters.] 
  In mathematical models of
  DS, it is customary to distinguish dynamical \emph{state variables}
  from \emph{control variables}. Both appear as arguments in the
  defining function of iterated maps and differential equations (and
  other formalisms), as in the generic ODE
  $\dot{\mathbf{x}} = f(\mathbf{x}, \mathbf{a})$ where $\mathbf{x}$ is
  the state vector and $\mathbf{a}$ denotes the vector of control
  parameters. The idea is that the latter are ``fixed'' or ``given''
  and are not affected by the system state update operators. However,
  this role distribution dynamical vs.\ fixed is not always
  clear-cut. Often one considers scenarios where the control
  parameters are subject to slow changes, for instance induced by
  top-down regulatory input in hierarchical neural processing
  architectures.

 \stepcounter{tdimension}\label{absRelScales}
\item[\Alph{tdimension} Absolute and relative timescales.] Some systems, formal or
  physical, can be ``run'' faster or slower and the speed is the only
  thing that changes. Examples are systems defined by differential
  equations, whose ``velocity'' can be set by time constants; or digital
  microprocessor systems whose clock cycle duration can be varied. To
  describe this ``velocity'' one needs an absolute reference time which
  can be formal (as in ODE systems) or physically ``real'', as in
  physical microprocessors. Absolute timescales are important for
  computing systems that are interacting with their input/output
  environment ``in realtime''.

  Many of the fascinating properties of complex dynamical systems
  arise not from its absolute ``velocity'' but from the fact subsets
  of dynamical variables, or subsystems, evolve faster or slower than
  other subsystems.  Temporal multiscale properties can also be
  attributed to the dynamics of a single variable: In a
  single-variable timeseries one may identify a spectrum of short- and
  longrange ``correlations'', or ``memory traces'', or statistical
  dependencies, etc. \emph{Multiscale dynamics} are a general and
  maybe essential characteristic of \emph{complex} systems, although
  this concept has no single, commonly agreed definition.

  It is remarkable that there seems to be no good word for the
  ``velocity'' of a dynamical system, which is why we put this word in
  quotes. In speaking about ``velocity'', one says ``the system is
  slow'' or ``the fast subsystem'', but no-one says ``the speed (or
  velocity) of the system is high''.
 \stepcounter{tdimension} \label{measuring}
\item[\Alph{tdimension} Measuring time.] For a physicist or signal processing engineer,
  time is given by nature and invariably denoted by
  $t$. Mathematicians do not care about real time and speak of ``unit
  time steps'' or ``the unit time interval'', an arbitrary convention
  to associate the unit inverval on the real line as a reference to
  quantify time. When a mathematician talks with a signal processing
  engineer, the former tends to be puzzled (if not disturbed) by the
  fact that the latter keeps talking about ``seconds'', a word that
  one will not find in mathematical textbooks on dynamical
  systems. Theoretical computer scientists ignore the aspect of
  temporal duration entirely. The formal models of computing automata
  only know of ``state update steps'', where the only aspect that is
  left over from physical time $t$ or unit time $[0, 1]$ is
  \emph{serial order} and causation: the next state comes after the
  previous and the latter determines the former. A most interesting
  challenge with regards to measuring time arises for computational
  neuroscientists when they want to explain how a brain can
  ``estimate'' or ``experience'' time. What mechanism in neural
  dynamics can enable a subject to estimate the presentation duration
  of a stimulus? Proposed answers include the use of neural delay
  lines, neural reference oscillators that function as clocks, or
  stimulus-duration-characteristic patterns in high-dimensional neural
  transients. We find that a general theoretical treatment of how
  ``clocks'' or ``time-meters'' can be defined in dynamical systems
  would be a rewarding subject of study.   
 \stepcounter{tdimension}\label{collectiveDerived}
\item[\Alph{tdimension} Collective and derived variables.] In statistical physics,
  neural field models, population dynamics and many other domains
  where one investigates systems made from large numbers of
  interacting small subsystems or ``particles'', one often describes
  the global dynamics of the ``population'' or ``ensemble'' through
  derived collective variables. Their timescale is typically
  \emph{slower} than the native, local timescales of the interacting
  subsystems; and one typically tries to capture the global dynamics
  with a \emph{small number} of such collective variables. Slowness is
  here connected with \emph{dimension reduction},
  \emph{simplification} or \emph{abstraction}. 
 \stepcounter{tdimension}\label{subSuperSampling}
\item[\Alph{tdimension} Sub- and supersampling.] In discrete-time models of dynamical
  systems one can create ``speedups'' by subsampling and ``slowdowns''
  by supersampling / interpolation. This however makes sense only for
  discrete-time models that can be understood as sampled versions of a
  continuous-time process. It makes no sense to supersample,
  for instance, the state sequence of a Turing machine.

   \stepcounter{tdimension}\label{slowingByDiscretization}
 \item[\Alph{tdimension} Slowing-down by discretization.] When a
   real-valued timeseries is discretized by binning, or a fine-grained
   discrete-valued timeseries is further simplified by coarsening,
   high-frequency detail (which can be regarded as fast-timescale
   information) gets lost in cases where there are oscillatory
   fluctations within bins in the original timeseries.  Thus,
   discretization or binning procedures may cut the spectrum of
   effective timescales.

   \stepcounter{tdimension}\label{freqFiltering}
 \item[\Alph{tdimension} Frequency filtering.] Applying frequency
   filters to trajectories deletes dynamical components on the
   timescales corresponding to the cancelled frequencies. In the
   special case of low-pass (smoothing) filters, fast timescale
   information is lost. In the special case of high-pass (baseline
   normalization) filters the opposite effect is achieved. 
   
 \stepcounter{tdimension}\label{whatIsMoment}
\item[\Alph{tdimension} What is a ``moment''?] We are used to think of a timeline as an
  ordered sequence of \emph{time points} - let us call them
  ``moments''. Even when one leaves out the complications of
  relativity theory, seeing time as a succession of zero-time moments
  is not always the most helpful view. Cognitive neuroscientists tell
  us that the subjective experience of ``now'' in some ways integrates
  over several milliseconds. When neuroscientists try to detect or
  define ``synchrony'' in neural spike patterns, they must soften the
  mathematical notion of point-sharp co-temporality to short
  intervals. Signal processing engineers would sometimes like to get
  rid of delays in their equations but can't. Abstractly speaking, in
  high-dimensional dynamical systems with nonzero-length signal travel
  pathways, relevant information-carrying ``patterns'' arise not
  instantaneously but need some minimial duration to realize
  themselves. Such observations suggest that in hierachically
  structured complex dynamical systems, a hierarchy of
  ``nowness-windows'' might be an appropriate concept, with
  short-duration ``moments'' defined for small, local subsystems and
  increasingly longer-duration ``moments'' as one goes up in the
  subsystem hierarchy. 
\stepcounter{tdimension}\label{homeostasis}
\item[\Alph{tdimension} Homeostasis, stability, robustness.]
  Biological organisms, brains, non-digital microchips made from
  unconventional materials, and many other computing or cognizing
  systems must preserve their functionality in the
  presence of external perturbations, change of environment, aging,
  parameter drift and other challenges. They do so through a wide
  spectrum of stabilization mechanisms which exploit, for example,
  redundancies, attractor-like phenomena, stabilizing feedback
  control, adaptation and learning, or robust network topologies for
  system architectures. A common denominator in this diversity of
  mechanisms is that they aim to ensure that vital system
  variables stay within a (narrow) viability window, often by
  attempting to stabilize them close to an optimal value. This has a
  twofold aspect of slowness. First, change rates of variables that
  are being stabilized are slow (when the stabilization is
  successful). Second, these critical variables must be stabilized
  through long timespans --- vital variables through the entire system
  lifetime.   
\stepcounter{tdimension}\label{nonstationarityModes}
\item[\Alph{tdimension} Nonstationarity and mode hierarchies.]
  Computing systems exhibit nonstationary dynamics, be it because they
  are input-driven or because they ``learn'' or because they execute a
  sequence of subprograms. System trajectories (timeseries) resulting
  from nonstationary dynamics can be qualitatively or quantitatively
  described through temporal hierarchies of \emph{dynamical
    modes}. For instance, a neuronal spike train can be characterized
  on a very short timescale by an interspike interval, on a short
  timescale by burst modes, on a longer timescale by locally averaged
  firing frequency, and on a very long timescale by asymptotic
  measures.

  In information processing systems, one may find ways to characterize
  what the current mode ``represents''. For instance, the neural
  activity trajectories in a speech-processing brain might be
  described as ``encoding'' or ``representing'' linguistic phonemes,
  syllables, words, phrases, sentences, texts.

  There exists no unique, general mathematical characterization of
  modes. Modes of an evolving DS might, for instance, be characterized
  in terms of frequency spectra, signal shapes, signal energy,
  attractor structures, degrees of chaoticity, or regions of the
  system's state space, to name but a few.  Describing temporal
  multiscale dynamics is very much the same task as characterizing
  modes, and there seems to be an unlimited repertoire of options.

  \stepcounter{tdimension}\label{hierarchicalArchitectures}
\item[\Alph{tdimension} Hierarchical architectures.] The human brain,
  most autonomous robot control systems, and many multiscale signal
  processing and control systems are hierarchically structured. The
  ``bottom'' layers are in direct contact with incoming signals and
  generate output signals, while ``higher'' processing layers carry
  out increasingly ``cognitive'' tasks based on increasingly
  abstracted and condensed representations of the information
  contained in the input signals.

  We find it a wide-spread, even paradigmatic view that higher levels
  operate on slower timescales than lower levels. This view is
  supported by evidence from biological brains, and guides the design
  of artificial signal processing and control systems that have to
  cope with temporal multiscale data. It also agrees with the view of
  classical AI, where \emph{action planning} architectures generate
  goals and subgoals, plans and subplans, procedures and subprocedures
  in a nested way, where higher nesting levels are taken care of by
  higher processing layers.

  A formidable challenge arises for formal modelers and concrete
  system developers (in computational neuroscience, machine learning
  and robotics). It concerns the nature of ``top-down'' influences: in
  what sense, and by which concrete mechanisms, do higher layers
  influence the processing on lower layers? Should this influencing be
  understood and realized as attention, prediction, context setting,
  or modulation? Many questions, both conceptual and
  algorithmic/mathematical, still are open. 

  \stepcounter{tdimension}\label{grammarsMemory}
\item[\Alph{tdimension} Characterizing multiscale dynamics from left
  to right and from the side.] In symbolic
  dynamics and theoretical computer science, a theme related to
  multi-timescale dynamics is infinite-length symbol sequences. They
  can be characterized by automata models, where some type of
  automaton generates the sequence ``from left to right'' --- that is,
  the sequence is seen as the trajectory of a dynamical system. But
  such sequences are also described and analyzed as being the fixed
  points of applying grammar rules. This method of characterizing the
  structure of an infinite sequence is a-temporal but directly yields
  a transparent account of its multiscale, hierarchical
  structure. Research to connect these two views has only recently
  started.  It seems likely (even
  obvious) that multiscale properties of DS trajectories are related
  to memory mechanisms that are effective in the generating DS.

  In theoretical modeling of timeseries data (in theoretical physics and
  economics in particular), stochastic dynamics with \emph{long
    memories} are discussed in terms of the shape of the corresponding
  power spectrum. One speaks of \emph{fat} or \emph{heavy tailed} or
  $1/f$ power laws. Such long-memory behavior
  is associated with \emph{self-induced criticality} or \emph{edge of
    chaos} conditions, and is often claimed as a characteristic of
  complex natural processes, for instance in economics, neural
  dynamics, or speech.

  Theoretical computer science offers a canonical repertoire of
  methods to specify automata with increasing memory capacities (from
  finite-state autonomata through a variety of stack automata to
  Turing machines), and how they relate to an equivalent grammar. It
  would be interesting to investigate how such memory mechanism
  hierarchies of symbolic automata and their grammars can be
  transfered into the domain of continuous-time, continuous-value DS
  and the power spectrum proporties of their trajectories.

  \stepcounter{tdimension}\label{warping}
\item[\Alph{tdimension} Time warping.] In real-life timeseries one
  frequently finds local speed-ups or slow-downs, for example a
  speaker stretching out the pronounciation of a vowel for emphasis. A
  related effect occurs when different realizations of a signal are
  originating from slower or faster generators, for example from
  slower or faster speakers. Biological brains can, within limits,
  compensate for such \emph{time warping} in inputs. For artificial
  temporal pattern recognition systems --- often recurrent neural
  networks (RNNs) --- this poses serious challenges.

\stepcounter{tdimension} \label{onlineRealtime}
\item[\Alph{tdimension} Online and real-time processing. ]  Many
  applications of signal processing and control systems must generate
  output responses to incoming data streams without or only minimal
  delay. This is the generic case for control systems, but also for
  many other applications, for instance speech-to-speech translation,
  medical cardiographic monitoring.  One speaks of \emph{online} or
  \emph{real-time} processing. One may make a fine distinction (not
  always observed) between these two concepts.

  In online processing, the signal processing system is
  ``entrained'' to the driving input stream. Its internal states
  directly ``synchronize'' with the input, where ``synchronizing'' is
  understood in a generalized way that includes nonlinear
  transformations and memory effects. The processing dynamical system
  can appropriately be mathematically regarded as a dynamical
  system. Analog signal processing devices and RNNs are prototypical
  examples.

  In real-time processing --- a natively digital-computing notion ---
  the algorithmics of the system-internal processing is decoupled from
  the input. The input signal stream is sampled and buffered,
  processing subtasks are identified and solved by algorithms which
  must run fast enough to deliver results within predefined time
  limits. On universal computers this may require the use of an
  underlying real-time operating system. 
  
\stepcounter{tdimension}\label{timeComplexity}
\item[\Alph{tdimension} Time complexity classes.] In theoretical CS,
  input-output tasks that can be algorithmically solved (i.e.,
  \emph{computable} tasks) are ordered into a hierarchy of \emph{time
    complexity}. In theoretical CS, input arguments are always
  formatted as finite-length symbol strings (``words''). The runtime
  of an algorithm is measured by the number of machine update steps
  (concretely, clock cycles) needed from the presentation of the input
  word until the output word has been generated. To define a
  complexity class, the runtime is related to the length $|w|$ of the
  input word $w$. For example, the class $P$ of polynomially
  computable tasks comprises all tasks for which some algorithm (or
  deterministic machine) and some polynomial $p$ exist, such that the
  algorithm terminates within $p(|w|)$ update steps, for all input
  words $w$.

  We remark that the concept of time complexity is tied to
  understanding ``computing'' as ``running a Turing machine from
  presenting an input word until it terminates with an output
  word''. This concept of time complexity cannot be naturally
  transfered to online processing tasks.
  
\stepcounter{tdimension}\label{TSnamings}
\item[\Alph{tdimension} Different names for different timescales.]
  Biological brains exhibit dynamical processes on many timescales,
  and different processes affect different physical elements in brains
  in different ways. This leads to a entangled maze of dynamical
  phenomena in which it is hard to not get lost. A coarse
  orientation is provided by the conceptual sequence \emph{inference}
  $\to$  \emph{adaptation} $\to$  \emph{learning} $\to$
  \emph{development} $\to$  \emph{evolution}. These terms denote 
  denote bundles of dynamical phenomena which manifest themselves on
  increasingly long timescales. None of them has a precise definition,
  but all of them are used in computational science, cognitive science
  and neural-networks based machine learning with more or less
  similar semantic intuitions:
  \begin{itemize}
  \item \emph{Inference} processes refer to the fast operations of
    sensor processing, motor control and ``reasoning'' which
    do not essentially rely on structural or parametric changes of the
    neural processing system, using the system ``as is''. In machine
    learning one often speaks of ``inference'' when a ready-trained
    neural network (or other ML model) is used to process task
    instances for which it has been trained. 
  \item \emph{Adaptation} is a particularly broad and vague concept. A
    common denominator of its uses seems to be that adaptation works
    on slower timescales than inference, and is in principle
    reversible. It often describes processes when a cognitive / neural
    system re-calibrates, or re-focusses itself when the environmental
    context of operation changes. In formal models, adaptation
    processes often are expressed through changes of control
    parameters in neural subsystems, induced by ``top down''
    regulatory mechanisms or subsystem-inherent homeostatic
    self-stabilization mechanisms. While this seems to us the most
    common intuition connected to the word ``adaptation'', it is also
    used in a much more generalized way to denote any change of any
    sort of system (from a single synapse to a biological population
    in an ecological niche) that improves the system's ``performance''
    or ``viability''. In those cases, adaptation is not usually
    reversible. 
  \item \emph{Learning} refers to processes which expand the
    functionality of a cognitive system on the basis of
    experience. Learning processes are usually considered irreversible
    (``forgetting'' are processes in their own right which cannot be
    understood as time-reversed learning). Learning processes are
    commonly associated with irreversible changes in system parameters
    --- in neural networks typically ``synaptic weights''. Structural
    changes (like deletion of neural connections or adding neurons to
    a network) may also result from learning, though this aspect seems
    less central to the ``learning'' concept than mere parametric
    change.  
  \item \emph{Development} is a notion which is much more common in
    the cognitive and neurosciences than in machine learning. It
    refers to the life-long history of an individual, autonomous
    cognitive system (animal, human, or generalized
    ``agent''). The development history is often segmented into life
    periods like pre-natal development, stages of infancy, youth,
    adolescence, old age which are in turn associated with specific
    structure-changing processes in the agent's brain. We foresee that
    developmental change will also become an important theme in
    neuromorphic computing systems based on non-digital hardware which
    cannot be ``programmed'' and whose physical substrate is subject
    to aging. 
  \item \emph{Evolution} is the longest-timescale item in our list
    of process categories. It describes the adaptive change of entire
    populations, across generations, to fit a (possibly changing)
    environmental ``niche''.     
  \end{itemize}

  Mathematical models of cognitive systems describe inference and
  adaptations processes (typically) through changes in the values of
  system variables (dynamical state variables and/or control
  parameters). The system equations do not structurally change. In
  contrast, models of development and evolutionary processes must
  account for structural changes in the system equations. Formal tools
  for \emph{effecting} and \emph{simulating} structural change in
  system equations exist in the form of genetic / evolutionary
  algorithms. However, mathematical theories that can be used to
  \emph{characterize} and \emph{analyse} structural change \emph{in
    qualitative terms} are scarce, heuristic, and generally still
  under-developed. We find that certain tools in mathematical logic
  (``non-monotonic logic'') come closest. However, these formalisms
  are not connected yet to dynamical systems modeling.

  \stepcounter{tdimension}\label{philosophy}
\item[\Alph{tdimension} Philosophy of time.] Time is a fundamental
  quality of human experience, and philosophical inquiries have
  approached this theme from many angles. This lies outside our
  competences and we only list some of the aspects of time that have
  been investigated by philosophers (gleaned from \cite{Callender11}).
  \begin{itemize}
  \item \emph{Time and metaphysics}. What are the ontological realities
    (``presentalist'', ``possibilist'', ``eternalist'') of the
    past, the present, and the future? Is time continuous or discrete?
    
  \item \emph{The direction of time.} What is the difference between
    the past and the future? Is the arrow of time inherent in time, an
    effect of causality, or of thermodynamical laws?
  \item \emph{Time, ethics and experience.} Themes include: the
    subjective ``now''; memory, anticipation, decisions and free will;
    development of time concepts in children; benefit and harm in the
    past and the future.
  \item \emph{Time in physics.} Are (which) physical laws time
    reversible? How is time related to space? How is time understood
    in relativity theory and quantum theory? What are clocks? How is
    time affected by the uncertainty principle? Is there time at all?
 
    \end{itemize}

  \end{description}

  This listing of aspects of time's ways of flowing faster or slower,
  or of our ways to observe a system for shorter or longer durations,
  makes it clear that a systematic, unified account of ``timescales''
  is out of reach. In order to give instructive initial input to the
  MemScales project, the best we can currently do is to compile a
  ``tourist guide''-like collection of concrete empirical findings,
  mathematical models and theories, and machine learning approaches
  which have a bearing on some of the listed dimensions of
  ``timescales''. We coarsely sort these collection items into three
  sections \textbf{Brains}, \textbf{Mathematics} and
  \textbf{Computing}, which is somewhat arbitrary since many lines of
  research cross-connect these areas. Our survey will be all but
  complete: firstly because we largely omit entire domains of science
  (in particular biology, physics, psychology and philosophy), and
  secondly because even in the three domains that we did explore (the
  wider neurosciences, mathematics and computer science / machine
  learning), our bounded expertise and the breadth of the subject put
  limits on what we could effectively cover. At the end of each
  section we list themes that we know should be included in future
  extensions of such a survey.


  \section{Brains}\label{secBrains}

  This section collects perspectives of research, empirical findings
  and models from the wider neurosciences including (some of)
  cognitive science.
  
  The title of the first four subsections are the consequence of
  frameworking towards a theory of neuromorphic signal processing,
  which we hope to work out more fully and more systematically in our
  future work in Mem-Scales. A
  metaphor to motivate our strategy is that it is incredibly difficult
  to solve a Rubik’s cube by just focusing on one side at a time. In
  that analogy, it is even undefined which parts of the cube
  correspond to an open problem. For in some definitions, there
  already are reasonable solutions to so-called open problems like the
  binding problem \citep{skarda1999perceptual}, stability-plasticity
  dilemma \citep{shouval2002unified}, and systems memory consolidation
  \citep{van2012schema}. But these solutions do not bind together to a
  comprehensive theory. The organization of material in our four
  subsections below arises from distinguishing two axes of
  discussing neural timescales, an individual (neuron) --- recurrent
  axis and a dynamical processing --- plastic adaption axis.

\subsection{Single-neuron processing}\label{secSingleNeuronProc}

A biological neuron has multiple timescales of phenomena due to
voltage-gated \citep{doyle1998structure} and ligand-gated
\citep{katz1971quantal} ion channels \citep{ranjan2011channelpedia},
spatiotemporal filtering across dendritic cables \citep{rall2009rall},
hierarchical synaptic-dendritic-membrane-somatic processing
\citep{gao2018dendritic} and biochemical pathways involving multiple
chemical compounds \citep{bray1995protein, barkai1997robustness,
  bargmann2006chemosensation}. Thus, a single neuron has enormous
capacity for signal processing, much better than the McCulloch-Pitts
and LSTM units in presently widespread artificial neural networks.

A special primitive for spatiotemporal processing at the dendritic
level is coincidence detection. It can explain concentration-invariant
signal recognition, for example in olfactory
\citep{hopfield1995pattern} networks. Chaining of multi-timescale
transient units with a coincidence detector results in transient
synchrony \citep{hopfield2001moment} and can explain uniform
time-warping invariant signal recognition.

A pioneering model of temporal processing at the membrane level is due
to Hodgkin and Huxley \citep{hodgkin1952quantitative}, which considers
the membrane potential and ion channel activation numbers as a
coupled system of nonlinear  ODEs. A generalization of the
Hodgkin-Huxley model with multiple ion channels whose conductances are
nonlinear and modulatable at multiple timescales is now the gold
standard for modelling the membrane dynamics of a neuron. Note that if
the ion channel activation numbers do not have any inter-neuronal
immediate effects, then just modelling the membrane potential is
sufficient for a complete neurodynamical understanding. For example,
\cite{izhikevich2004model} showed that a reduced
2-dimensional threshold-reset ODE system is sufficient to explain a
possible set of 20 kinds of temporal processing in cortical neurons
including tonic spiking, phasic spiking, spike bursting, spike
latency, subthreshold oscillations, rebound, bistability, and spike
frequency adaptation.

\subsection{Recurrent processing}\label{secRecProc}
Here we will focus on the workings of recurrent neural network,
ignoring plasticity. Functionally, a worm brain can be understood to
operate in sensory-inter-command-motor layers
\citep{gray2005circuit}. The human brain is similar but more
complicated \citep{eliasmith2012large}, i.e. the command-layer is
split into functional regions performing action selection and motor
processing, and the inter-layer is split into functional regions
performing information encoding, transform calculation, reward
evaluation, working memory and information decoding. 

Three noteworthy observations arise from the study of recurrent
processing with multiple timescales. Firstly, there often exists a
\emph{behavioural hierarchy} \citep{davis1979behavioural} resulting in a
‘singleness of action’ where a long timescale state controls shorter
timescales. For example, the mating state of stickleback fish
activates a stereotypical dance movement \citep{tinbergen1951study},
honeybees in a communicative state employ a waggle dance routine
\citep{von1967dance}. Secondly, the behavioural hierarchy can be \emph{deep},
as in a reproductive instinct that activates sub-behaviours such as
nest-building or defensive fighting. Experiments have shown that a
three level hierarchy explains worm locomotion both behaviorally and
in neuroanatomy \citep{kaplan2020nested}. Lastly, there need not
always be an equivalence between anatomical and behavioral hierarchy,
for example chains of neurons can generate birdsongs
\citep{long2010support}.

\subsection{Neuron-neuron plasticity}
Here we will consider the form of synaptic plasticity as postulated by
Hebb \citep{hebb1949organization}, where any change in the synaptic
weight from one neuron to another neuron, is only based on signals due
to the activity of the two neurons i.e. deterministic bi-terminal
interactions. Networks with Hebbian plasticity, with or without memory
of the neuronal activity (an extreme case is a strict "spike-time"
dependence \citep{caporale2008spike}), are theoretically capable of
signal processing primitives such as principal component analysis
\citep{oja1982simplified}, self-organizing maps
\citep{kohonen1982self}, and independent component analysis
\citep{jutten1991blind}. So, at a network level, Hebbian learning can
be much deeper than the popular maxim of “cells that fire together,
wire together”.  Also, Hebbian-like learning is possible within a
single cell \citep{fernando2009molecular} if they contain motifs of
chemical cycles where the concentration of different chemical species
(such as in gene regulatory networks or phosphorylation cycles) can
mimic the functionality of synaptic weights (slow-varying control
parameters) and action potentials (fast-varying state variables).

Among the gamut of possible Hebbian plasticity rules, the most
noteworthy is the Bienenstock-Cooper-Munro (BCM) model
\citep{bienenstock1982theory} because it has been experimentally
justified \citep{cooper2012bcm}. The BCM model has the rate of change
of the synaptic weight equal to  a fast timescale times the
correlations in the neuron-neuron activity times a saliency factor
equal to a mean-deviation of the postsynaptic activity, minus a slow
timescale times the synaptic weight. Thus, the BCM model has an
increased rate of forgetting on introducing uncorrelated noise, can
converge in whitened environments by means of higher-order statistics,
can learn direction sensitivity without relying on neuroanatomical
asymmetry, and can have a single neuron to be both directionally and
orientationally sensitive by learning on video stimuli.

Also noteworthy is that a biophysical model of bidirectional synaptic
plasticity \citep{shouval2002unified} can be phenomenologically
reduced to a voltage-based STDP \citep{clopath2010connectivity}, which
under certain input conditions is equivalent to the BCM
model. Experimental measurements of STDP on the visual cortex,
somatosensory cortex and hippocampus could be fit to the
phenomenological model and distinct timescales were extracted.

\subsection{Recurrent plasticity}
We can look at recurrent plasticity as having two sides :
deterministic effects and non-deterministic effects.

All deterministic effects that are beyond bi-terminal interactions can
be subsumed under the banner of neo-Hebbian plasticity
\citep{gerstner2018eligibility}, including neuromodulated STDP
\citep{fremaux2016neuromodulated} due to some form of reward or
punishment (leading to the concept of an eligibility trace and
three-factor rules) and heterosynaptic plasticity due to some
conservation law \citep{oh2015heterosynaptic} such as the spatial
conservation of the total synaptic weight \citep{von1973self} or the
normalization of the synaptic weights \citep{hyvarinen1998independent}
to energetically sustainable levels
\citep{walker2006sleep,kandel2014molecular}. Note that neo-Hebbian
plasticity combined with a suitable inter-layer (that is capable of
generating rewards internally for congruent or novel information) is
sufficient for effective systems memory consolidation
\citep{van2012schema}, but of course in reality non-determinism will
also play a supplementary role as discussed below.

All non-deterministic effects (including bi-terminal interactions) can
be subsumed under the banner of neural Darwinism
\citep{edelman1993neural}, also known as neuroevolution
\citep{stanley2019designing}. It is plasticity that is based on the
principle of selection upon variation, and hence is biased towards
generating a hierarchical organization. Experimental evidence supports
a hierarchical organization in the basal ganglia to generate action
sequences \citep{jin2015shaping}, so it can play an important role in
learning procedural memory. Of course, at some level genetics can also
enforce a hierarchy \citep{felleman1991distributed}, but there is a
reason to believe that neural Darwinism plays a major role given that
hierarchies in the brain themselves are adaptive. For example, even
people with cerebellar agenesis learn to walk
\citep{boyd2010cerebellar} and spoken language perception colonizes
the visual cortex in blind children
\citep{bedny2015visual,lane2015visual}. Also, neural Darwinism can
work for hard problems like nonlinear blind source separation for
which deterministic and global optimization methods like slow-feature
analysis end up failing in high dimensions \citep{WiskottSejnowski02}.

\subsection{Some topics not covered}
We list a number of themes that would warrant a closer inspection but
for which we lacked time or expertise on this occasion.  Surely there
is an endless list of such themes. Nevertheless, with deeper thought
or moments of serendipity, we should work towards an ideal where newer
themes are assimilated or accommodated into older themes (the
Piagetian pun is intended \citep{piaget1952origins}).

  \begin{itemize}
  \item Neural clock circuits and entrainment of neural dynamics to
    clock signals. 
  \item Variable binding through theta-wave phase synchronization.
  \item Hierarchies of memory mechanisms --- a large research
    field in the cognitive and neurosciences which would need a
    separate, extensive treatment. Surveys are given, for example, by
    \cite{Durstewitzetal00} or
    \cite{FusiWang16}.
  \item The role of cerebellar processing in timing fine-control.
  \item Experimental demonstrations of different time constants in
    cortical processing \citep{Bernacchiaetal11}.
  \item Perception of temporal patterns \citep{LargePalmer02}.
  \item Stages in ontogenetic development.
  \end{itemize}

\section{Mathematics}\label{secMath}

In this section we collect ``pure'' mathematical themes and formal
modeling methods from dynamical sytems theory and some areas of
theoretical physics. Topics in formal logic are treated in Section
\ref{secComp}.

\subsection{Singular perturbation theory}\label{}
Arguably the most popular mathematical approach to studying multiple
timescale dynamics has been via \textit{singular perturbation theory}
(SPT) of systems of ODEs. Intuitively, this theory studies
perturbations with small parameters where the dynamics cannot
straightforwardly be approximated by the limiting case where the
parameters vanish \citep{OMalley1991, Verhulst2005}. A geometric
approach to singular perturbation theory (GSPT) was first set up by
the works of A.N. Tikhonov and those of N. Levinson
\citep{Vasileva1967, Kaper1999}, later worked out in more detail
by N. Fenichel \citep{Fenichel1979}. This geometric approach
formalizes interpretations of certain singularly perturbed systems as
`slow-fast' systems, where some variables operate on a relatively fast
timescale compared to other slower evolving variables. An enormous
amount of research has been done on these slow-fast systems, as they
are relevant for the mathematical description of many processes in the
life sciences. An extensive modern overview of mathematical theory on
slow-fast ODE systems is given by \cite{Kuehn2015}. In particular,
slow-fast systems have become a big research topic in mathematical
neuroscience, see for example \cite{RubinTerman2002, Izhikevich2007,
  Ermentrout2010, Pusuluri2020}. Another notable application of SPT is
for control science, where the classical text is \cite{Kokotovic1999}.

To illustrate the mathematical approach to slow-fast ODEs, consider
the two-dimensional system
\begin{align}
\begin{split}\label{eq1}
\tau_1 \frac{dx}{dT} &= f(x,y),\\ 
\tau_2 \frac{dy}{dT} &= g(x,y),
\end{split}
\end{align}
with $f$ and $g$ some possibly non-linear functions. We assume $\tau_1, \tau_2 > 0$ to represent the intrinsic timescales of respectively the $x$ and $y$ variables. Now define a new parameter $\epsilon = \tau_1 / \tau_2$. Then system \eqref{eq1} can be transformed both into
\begin{align}
\begin{split}\label{eq2}
\epsilon \frac{dx}{ds} &= f(x,y),\\ 
 \frac{dy}{ds} &= g(x,y),
\end{split}
\end{align}
and into 
\begin{align}
\begin{split}\label{eq3}
\frac{dx}{dt} &= f(x,y),\\ 
 \frac{dy}{dt} &= \epsilon g(x,y),
\end{split}
\end{align}
via reparameterizations of time $T=s \cdot \tau_2$ and
$T = t \cdot \tau_1$ respectively. As long as $\epsilon > 0$, systems
\eqref{eq2} and \eqref{eq3} can be considered equivalent. Suppose that
$\tau_1 \ll \tau_2$; intuitively this means the $x$-variable operates
on a much faster timescale than the relatively slow $y$-variable. In
that case $0 < \epsilon \ll 1$, and we might consider how systems
\eqref{eq2} and \eqref{eq3} for $\epsilon>0$ behave when approaching
the singular limit $\epsilon \rightarrow 0$. We may now also call $s$
the slow timescale, and $t$ the fast timescale. For $\epsilon=0$, it
is important to remark that systems \eqref{eq2} and \eqref{eq3} are
not equivalent anymore. The case $\epsilon=0$ for system \eqref{eq2}
is also referred to as the \textit{slow subsystem} or \textit{reduced
  problem}, while the case $\epsilon=0$ for system \eqref{eq3} may be
called the \textit{fast subsystem} or \textit{layer problem}.

Intuitively, for $0 < \epsilon \ll 1$ the layer problem approximately
describes the dynamics of the system on a short timescale, where the
slow variable $y$ can be approximated by a constant. Therefore, $y$
can be interpreted as a bifurcation parameter of the fast
subsytem. The reduced problem describes the dynamics at $\epsilon=0$
of the slow variable on a one-dimensional manifold $C_0$, also called
the \textit{critical manifold}, which is given by the zeros of
$f$. Observe that $C_0$ can alternatively be said to be given by the
equilibria of the fast subsystem \eqref{eq3}. Close to attracting
hyperbolic parts of $C_0$, the slow subsystem approximately describes
the dynamics of the system for $0 < \epsilon \ll 1$ on the longer
timescale represented by $\tau_2$. This is formalized by
\textit{Fenichel's Theorem}, see for example \cite{Fenichel1979,
  Kaper1999} or Chapter 3 of \cite{Kuehn2015}.

Orbits starting close to an attracting hyperbolic part of $C_0$ for
$0 < \epsilon \ll 1$ can be predicted to stay close to $C_0$ by
Fenichel's Theorem, approximating the flow of the reduced problem,
until nearing a point on $C_0$ where hyperbolicity is lost. This
happens at bifurcations, with respect to $y$, of the fast
subsystem. What happens after reaching such a bifurcation point,
requires careful analysis of the full system. One might find jumps
between attractors of the fast subsystem (stable equilibria in the
two-dimensional example under consideration here), as the system
converges towards the vicinity of another attracting hyperbolic part
of $C_0$. While this type of behavior often occurs in slow-fast ODE
models, immediate jumps between attractors of the fast subsystem
cannot be predicted in general from a decomposition of the full system
into fast and slow subsystems at $\epsilon=0$. Indeed, a peculiar type
of behavior might occur where the full system for some time
approximately follows a non-attracting hyperbolic part of $C_0$. This
phenomenon is known as a \textit{canard}, and for example plays a role
in the analysis of spike adding for models of bursting neurons in
mathematical neuroscience \citep{Terman1991, Linaro2012}.

The theory of slow-fast ODE systems has been extended to multiscale
stochastic differential equations (SDEs) incorporating noise, and more
generally to the context of random dynamical systems, see Chapter 15
of \cite{Kuehn2015}. Theory on slow-fast SDEs has for example been
applied to give a rigorous analysis of certain multiscale synaptic
plasticity models for neural networks in \cite{Galtier2012,
  Galtier2013}. Also, slow-fast systems of maps can be studied with
similar techniques, and have been applied to neuron models as well,
see for example \cite{MiraShilnikov2005} and \cite{Ibarz2011}.

\subsection{Delay equations}\label{secDelayEqns}
Multiple timescales can also be introduced in differential equations
via delays \citep{Yanchuk2017, Ruschel2020}. The simplest such delay
systems are modelled by
\begin{equation}\label{eq4}
\tau_L \dot{x}(t) = -x(t) + F(x(t-\tau_D)),
\end{equation}
where $\tau_L$ is the intrinsic time scale of the system, $F(x)$ is a
nonlinear function of $x$ and $\tau_D$ is the time delay. When
$\tau_D$ is large compared to $\tau_L$, it is known that these type of
systems can exhibit a host of interesting spatio-temporal dynamical
phenomena \citep{Yanchuk2017}. Intuitively, a comparatively large
delay introduces a slow timescale next to the fast intrinsic dynamics
of the system. Such delay systems with large delay have recently been
shown to be relevant for approaches to reservoir computing with
opto-electronic hardware \citep{Hart2019}. These opto-electronic delay
systems can be viewed as an alternative method for implementation of
high-dimensional neural networks. Space-time representations allow the
dynamics of a delay system with low spatial dimension to be
interpreted as spatio-temporal dynamics of spatially extended
systems. As such, delay systems have recently also been thought of as
useful for the study of complex dynamical behavior in large-scale
connected networks. Although delay systems were originally thought of
as similar to ring networks of identical neurons, \cite{Hart2019}
propose that delay systems can be used to implement networks with
arbitrary topologies.

By defining $\epsilon=\tau_L / \tau_D$, equation \eqref{eq4} can be
rewritten into the singularly perturbed delay equation
\[
\epsilon \dot{x}(t) = -x(t) + F(x(t-1)).
\]
Such type of such systems have been studied for example in
\cite{ChowMP1983} and \cite{IvanovSharkovsky1992}.

 \subsection{Some topics not covered}\label{secnotCoveredMath}

  We list a number of themes that would warrant a closer inspection
  but for which we lacked time or expertise on this occasion:
  \begin{itemize}  
  \item Critical slowdown of dynamics close to bifurcations
  \item Characterizing multiscale structure in infinite symbol sequences via fixed points of grammar rule applications
  \item Reaction-diffusion systems
  \item Variables of multi-dimensional iterative maps can be given
    differing update frequencies. Little (if any) formal mathematical
    theory seems to exist on this topic.
  \item Line attractors and their generalizations.
  \item Statistical physics modeling of collective phenomena and
    generalized synchronization, slaving principle \citep{Haken83}.
  \end{itemize}

\section{Computing}\label{secComp}

This section collects topics, techniques and models from computer
science --- machine learning and artificial neural networks in
particular. The ordering of subsections is arbitrary.

\subsection{Neuron models with time
  constants}\label{secdynamicalneurons}
In neural network (NN) architectures used in machine learning, a
variety (but not a very large variety) of formal/algorithmical neuron
models is used. The neuron models used in feedforward NNs always have
a-temporal state update equations of the kind
$x_i = f_i(\sum_j w_{ij} x_j + b)$, where the $x_j$ are the
activations of neurons feeding into neuron $i$ and $f_i$ is a (almost
always) monotonically growing ``activation function''. Time becomes a
relevant theme only in recurrent neural networks (RNNs). Besides the
a-temporal models $x_i = f_i(\sum_j w_{ij} x_j + b)$, which can also
be used in discrete-time RNNs (which then mathematically can be
regarded as implementing iterated maps), here we find a diversity of
neuron models that either are specified by ordinary differential
equations (ODEs) --- from the simple leaky integrator neuron
$c \dot{x}_i = - x_i + f_i(\sum_j w_{ij} x_j + b)$ with time constant
$c$ to LSTM units \citep{HochreiterSchmidhuber97} to multi-variable
circuit equations for use in analog VLSI neuromorphic microchips
\citep{Chiccaetal14}) --- or by discretized versions of such ODE
models, typcially using the elementary Euler approximation; or spiking
versions which include a discontinuous neural state reset
operation. All of these contain time constants. In complex neuron
models (often with a biological motivation), different time constants
can be set for different variables. For instance, slow synaptic
efficiency adaptation rules (``slow'' relative to the soma potential
dynamics) are crucial for creating dynamical memory traces in liquid
state machines \citep{Maassetal01}. When these neurons are
``executed'' in digital simulations, they can be made to ``run''
faster or slower over a wide range (limited only be numerical
stability conditions) compared to each other or to some reference
timescale. In analog neuromorphic hardware realizations however, these
time constants are fixed by physical givens and changing them is only
possible if the chip design allows one to access and ``set'' the
physical correlates of time constants (for instance voltages), and
 within limited ranges.

\subsection{Time and memory in digital
  computing}\label{secdigitalTime}
It is easy to \emph{simulate} multiscale temporal dynamics of
time-discretized ODE models on digital machines --- all one has to do
is to set different desired time constants in the various variable
equations.

It is not always easy to \emph{realize} multiscale dynamics of
time-discretized ODE models on digital machines when the computed
dynamics must match physical ``real-time'' in online signal processing
and control applications. The system state update equations must be
simple enough, or implemented cleverly enough, or parallelized enough,
to ensure that the digital processing needed to compute the next time
slice state takes at most as much time as the physical time allotted
for a sampling interval. This may become demanding in robot control
applications, in particular in compliant robots where the sampling
frequency must be high (order of 1000 Hz) in order to react fast
enough to sensor signals signifying effector impact.

In serial-data ``cognitive level'' AI / machine learning tasks like
text processing or visual gesture analysis, the processing algorithm
must have a working memory which has (at least) the power of a stack
memory. This is needed to process the hierarchically nested temporal
structure in ``grammatically'' organized input sequences. Realizing
such a stack memory is of course not a problem for digital computers
when they execute symbolic-AI programs. When RNNs that have been set
up or trained for such tasks are simulated on digital machines, the
hierarchical memory organization cannot be directly mapped to the
(easily available) physical stack memory mechanisms of the digital
computer. Instead, this memory functionality must become encoded and
realized in terms of the RNNs dynamics. One way to do so is to train
binary context-level switching neurons which can set the (single) RNN
into a temporal hierarchy of dynamical processing modes
\citep{PascanuJaeger10}.  

\subsection{Time complexity}\label{secTimeComplexity}
  In theoretical (symbolic-digital) computer science the concept of
  \emph{time complexity} refers to upper bounds on the maximal number
  of processing steps needed by a Turing machine to compute its result
  when it is started on a (any) input word of length $n$
  \citep{Jaeger19a} This leads to a classification scheme for the
  ``difficulty'' of computing problems. For instance, the time
  complexity class $P$ is the set of all input-output computing tasks
  such for a task $T \in P$ that there exists some Turing machine
  $\mathcal{M}$ and some polynomial function
  $p: \mathbb{N} \to \mathbb{N}$ such that $\mathcal{M}$ always
  terminates within $p(n)$ update steps. Some of the deepest unsolved
  problems in theoretical computer science (and indeed, in
  mathematics) concern such time complexity classes, in particular the
  famous $P =? NP$ problem \citep{Cook00}.

  This standard usage of the term ``time complexity'' is confined to
  characterizing the computational demands of evaluating
  \emph{functions} --- a Turing machine (and all other, equivalent
  mathematical definitions of an \emph{algorithm}, of which there are
  many) incorporates an input-word to output-word mapping. In the
  context of neuromorphic computing and recurrent neural networks, a
  dynamical systems interpretation of ``computing'' seems more
  adequate than a function-evaluation interpretation. Some models of
  ``computing'' have been proposed in theoretical computer science
  which account for continual online processing of unbounded-length,
  symbolic input streams, in particular \emph{interactive Turing
    machines} \citep{vanLeeuwenWiedermann01} and more recently
  \emph{stream automata} \citep{Endrullisetal20}. In followship of the
  traditional questions that are considered in classical complexity
  theory, this research aims at classifying continual input-output
  stream processing tasks into complexity classes. The adopted
  perspective on discussing such complexity classes is however still
  tightly tied to the classical, function-based concepts of time
  complexity, in that such automata are designed in a way that upon
  reading a new input symbol, they can ``detach'' from the input
  stream, do a possibly highly complex computation in traditional
  Turing machine fashion, and after this computation terminates,
  produce an output symbol (or not). Complexity class hierarchies
  investigated in such research typically include classes of continual
  serial input-output tasks which are inaccessible by physical
  machines --- \emph{super-Turing} tasks --- in that
  \emph{oracles} are invoked, that is, external additional input
  (outside the input data stream) is allowed which provides
  information that itself is not Turing-computable.

  There is also a body of research which aims at transfering concepts
  and methods from symbolic complexity theory to RNNs with real-valued
  activations and/or weights
  \citep{SiegelmannSontag94, SimaOrponen03}. One common theme and
  finding in this line is that a single infinite-precision real
  numbers allows one to encode an infinite amount of information,
  which gives rise to super-Turing computing powers (that is, symbolic
  input-output functions can be computed which no Turing machine can
  compute). However, there is no evidence that
  super-Turing performance cannot be physically realized due to noise
  and limited precision (observability) of physical state variables. 

\subsection{Logic formalisms for capturing
  time}\label{secLogicFormalisms}
A traditional topic in classical (symbolic) AI is ``reasoning about
action and change'', or ``reasoning about action and time'' (with
several conferences and workshops and a wealth of publications that
have these expressions in their titles). The objective of this
research is to extend the expressive powers of logic-based knowledge
representation and inference formalisms to facilitate the
representation of, and formal reasoning about, a cluster of themes
that includes action, change, planning, intentions, time, events,
causation and more. Such formalisms are algorithmically processed with
so-called \emph{theorem provers} (also called \emph{inference
  engines}). These are heuristic, discrete combinatorial search
algorithms whose processing steps are not interpreted as temporal
steps but as logical arguments. Time, timing, measuring time,
comparing durations, ordering events on a timeline etc.\ are objects
that are logically reasoned \emph{about}, in reasoning steps whose
ordering is conceived as logical, not temporal. More than four decades
of research have produced a rich body of representation formalisms. We
can only pinpoint a few examples. \cite{Allen91} is an early survey.
Fundamental figures of reasoning about time are captured by modal
operators in \emph{temporal logics} (also known as \emph{tense
  logics}) \citep{Garson14}. A related classical subfield of AI,
\emph{qualitative physics} \citep{Forbus88} (closely related:
\emph{naive physics}, \emph{qualitative reasoning}) explores
logic-based formalisms which capture the everyday reasoning of humans
about their mesoscale physical environment. A rather recent
development is hybrid logic / dynamical-systems formalisms to reason
about physical dynamical sytems \citep{Geuversetal10} in ways that
capture the measurement metrics of ``real'' continuous time. Such
formalisms are intended for formal verification of hybrid
physical-computational systems in systems engineering.

\subsection{Slow feature analysis}\label{secSFA}
Slow feature analyis (SFA), developed by L. Wiskott
\citep{WiskottSejnowski02, Wiskottetal08}, is a method to extract
features from timeseries data (in particular video streams) which are
defined by the fact that they change slowly. SFA has been used to
explain the functioning of feature detection cells in visual cortex
\citep{BerkesWiskott03} and hippocampal place cells
\citep{SchoenfeldWiskott15}. Interestingly (and possibly, limitingly),
the slow features found by SFA are functions of single input frames,
not --- as one might expect --- functions of input episodes that last
nonzero time. 

\subsection{Speed control in RNNs}\label{secslowRNNs}
Biological brains can generate and recognize instances of output
patterns which differ from each other only in speed (for instance,
generating or recognizing slow and fast hand-waving or dance or music
pattern generation). For a mathematical model of an RNN written in
ODEs with time constants, it would be straightforward to adjust the
processing speed in generation or recognition by scaling all time
constants with the same factor. But physical neural systems, whether
biological or in neuromorphic hardware, cannot scale all physical time
constants with a global factor. This leads to a very interesting
mathematical and biological (and algorithmical) question: how can the
qualitative dynamics of an RNN be speeded up / slowed down without a
globel time constant scaling? We are aware of two approaches which
both make use of the fact that when a RNN is excited by
different-speed versions of the same input pattern, the elicited
network states populate different regions of state space. By
characterizing the geometry of these different regions with (cheaply
computable) variables, and subesquently actively controlling these
variables by elementary linear controllers \citep{wyffelsetal13} or
conceptor filters \citep{Jaeger14}, speed variations up to a factor of
10 for pattern generating tasks have been achieved. 

\subsection{Space to time transformation}\label{secspace2time}
Even outside relativistic physics, space and time depend on each
other. In particular, travelling solitons and waves need time
proportional to travel distance. This may become exploitable for the
design of neural mechanisms for variable-speed pattern recognition and
generation. The idea is to encode the target pattern spatially on a
neural surface and let it be ``read'' by a travelling wave or soliton
whose travel speed is determined by a single or very few variables
that can be controlled physically or algorithmically. Neural field
theories of cognitive cortical processing, which are based on solitons
and waves, are worked out in some detail \citep{EngelsSchoener95,
  LinsSchoener14}, but as far as we can see, so far not with the aim
of explaining speed control.

\subsection{Behavior control hierarchies}\label{secControlHierarchies}
In robotics and intelligent agent modeling, the cognitive control of
action selection and motor control is typically organized in a
hierarchy of planning and controlling layers. Higher layers in such
hierarchies operate on slower timescales than lower layers. The lowest
layers Hierarchical agent ``architectures'' are so common and have
been proposed abundantly since 50 years, such that we can give only a
few, exemplary ad hoc pointers. Examples: In classical AI
architectures, such hierarchies are explained in terms of plan-subplan
hierarchies \citep{Saffiottietal95}. In control engineering,
hierarchical control architectures have become an explicit industry
standard \citep{Albus93}. In the archetype \emph{subsumption
  architecture} \citep{Brooks89} in behavior-based robotics,
higher-level ``behaviors'' can suppress lower ones. An influential
early model in neural networks / machine learning constitutes the
control hierarchy in a format of trainable hierarchical \emph{mixture
  of experts} where higher-level experts can \emph{gate} lower-level
experts. A similar structure, based on ODEs where higher-layer
behavior-controlling ODEs were run with slower time constants, powered
several winners in RoboCup world championships
\citep{JaegerChristaller97}. When mulit-layered RNNs are trained for
robot tasks, timescale-differentiated layers of control emerge
automatically \citep{YamashitaTani08}.

\subsection{Eligibility traces}\label{seceligibilityTraces}
In reinforcement learning (RL), a key submechanism is to represent and
compute \emph{eligibility traces} \citep{SuttonBarto98}. This refers
to a number of algorithmic methods to maintain a memory trace (with
weighted decay) of past actions (of a complete agent or a single
neuron, in the latter case the action being spike generation) and
input signals (sensor input to an agent or spike input to a neuron),
paired with information about the (estimated) utility of the action
history to receive reward. The setting of the decay rate determines
the memory horizon. Reinforcement learning can be expected to play a
large role in neuromorphic training.  Recently eligibility traces have
also become instrumental in designing neurally plausible (and hence
potentially implementable in physical spiking neuromorphic microchips)
approximate methods to emulate backpropagation learning in spiking
neural networks \citep{Bellecetal19}.

  \subsection{Time (un)warping}\label{secTimeWarping}
  In many temporal machine learning tasks, the incoming signal can be
  sometimes faster, sometimes slower. This can happen when signal
  sources change (for instance, there are slow and fast speakers), but
  it can also happen within a single instance of an input signal (for
  instance, when a speaker gets excited and speaks faster, or when
  his/her way of pronunciation has temporal ideosyncrasies like
  stretching vowels longer than average speakers). This is a problem
  for machine learning algorithms. In brute-force learning paradigms
  (deep learning in particular), such \emph{time warping} effects are
  caught by providing exhaustive training samples that cover all sorts
  of warping effects. A more training-data-economical approach is
  to send input signals through some time-unwarping preprocessing
  filter before feeding it to the RNN in training and exploitation,
  such that the RNN only has to cope with speed-normalized input
  signals. Another approach that we find the most elegant is to leave
  the input stream in its original time-warped version and adapt
  the processing speed of the RNN, speeding it up when the input
  signal slows down such that each RNN state update step (in
  discrete-time RNN models) or unit-time state evolution (in
  continuous-time RNN models) covers
  the same phenomenological change increment in the input stream
  \citep{Lukoseviciusetal06}.

  \subsection{Continual learning}\label{seccontinualLearning}
  \emph{Continual learning} refers to the machine learning challenge
  to make a neural network (feedforward or recurrent) learn a sequence
  of tasks, one after the other, such that when the next task is
  trained into the network, the new weight adaptations do not destroy
  what the network has previously learnt in other tasks. This
  \emph{catastrophic forgetting} (or \emph{catastrophic interference})
  problem has remained without a convincing solution since it was
  first acknowledged decades ago \citep{French03}. Only recently, a
  number of novel approaches in deep learning found effective
  algorithmic ways to de-fuse this problem. This is today a very
  active strand of research in deep learning, now called
  \emph{continual learning}, which has yielded a variety of effective
  algorithmic paradigms and a differentiated view on variants of the
  problem statement \citep{Parisietal19, Heetal19a}. The continual
  learning problem is closely related to the theme of \emph{transfer
    learning} (which concerns the generalization and carry-over of
  competences learnt on other tasks to a new task), the theme of
  \emph{federated learning} (which concerns the integration of
  learning progress made in the peripheral nodes in a network of
  decentralized local learners \citep{Kairouzetal19}), and the theme
  of \emph{meta-learning} (which concerns the learning of learning
  strategies).

  Continual learning is connected to timescale and memory topics in
  several ways. First, in some continual learning algorithms, weight
  changes in synapses that are deemed important for previously learnt
  tasks are discouraged, reducing (= slowing down) their adaptation
  rate. Second, the continual learning problem in its most demanding
  form poses itself on the longest possible, namely the
  \emph{lifelong} learning timescale. Third, some continual learning
  algorithms rely on generative memory replay of previously learnt
  tasks.

  \subsection{Dynamical memory in RNNs}\label{secMemInRNN}
  
  The simple linear readout which is typically used for training RNNs
  in the reservoir computing (RC) field can be used to define natural
  numerical measures for ``how much'' memory about previous input is
  preserved in the current network state. In its most basic format,
  the \emph{memory capacity} of a discrete-time ``echo state''
  reservoir network is measured by (i) feeding it with white noise
  input, (ii) training linear readout units $y_d$ by linear regression
  on the task to recover the input value $u(t-d)$ from $d$ steps
  before, (iii) adding all correlation coefficients between signals
  $y_d$ and $u(t-d)$ to get the desired measure
  \citep{Jaeger02b}. Note that the training of readout units here is
  not done to soleve a ``useful'' task but solely for quantifying an
  core characteristic of an RNN (or, for that matter, any other
  dynamical system). Note further that the ``memory'' which can be
  determined in this way is a purely dynamical short-term memory and
  involves no learning inside the RNN. This concept of memory capacity
  has become the anchor point for a (by now) extensive literature of
  mathematical analyses which explore memory in RNNs under aspects
  like the impact of noise \citep{Antoniketal18}, continuous time
  \citep{HermansSchrauwen10}, high-dimensional input
  \citep{HermansSchrauwen10a}, infinite-dimensional neural networks
  constructed by kernel methods \citep{HermansSchrauwen11}, different
  neuron models \citep{Buesingetal10}, or nonlinear readouts
  \cite{Grigoryevaetal16}, to name but a few. The literature is by now
  extensive and a systematic survey would be
  welcome. \cite{FradyKleykoSommer18} develop a classification scheme
  for dynamical memory tasks and measures which highlights the
  richness of phenomena and perspectives associated with dynamical
  memory in RNNs.

  The memory capacity of reservoir networks has become a standard
  metric to quantify or predict the ``goodness'' of reservoir networks
  for cognitive tasks in studies where different network architecures
  \citep{Straussetal12}, reservoir pre-training methods
  \citep{Schrauwenetal08}, or reservoir control parameter tuning are
  compared. The latter is often associated with investigations of
  reservoir performance ``close to the edge of chaos''
  \citep{LegensteinMaass07} (which in most cases
  we find an incorrect usage of terminology; correctly it should be
  ``close to the loss of the echo state property'').

  Measuring delayed-input to trained output correlations is not the
  only way is not the only way of quantifying the dynamical memory
  capacity of RNNs or general input-driven dynamical systems. If one
  adopts a probabilistic perspective, information-theoretic measures
  like the Fisher memory matrix \citep{GanguliHuhSompolinsky08} are
  more informative, albeit harder to estimate empirically.

  We point out that dynamical memory cast as state-based information
  carry-over from the past to the present, as discussed above, is not
  the same as \emph{working memory}. Working memory is a complex
  concept used in the cognitive and neurosciences for a spectrum of
  transient recall phenomena in animal and human remembering
  \citep{Baddeley03, BotvinickPlaut06, FusiWang16}. Working memory
  phenomena usually entail additional control mechanisms to encode and
  decode context information and insertion of knowledge stored in
  long-term memory. 

  Neither ``dynamical memory'', ``working memory'', nor ``short-term
  memory'' have generally shared, precise definitions and when one
  studies the literature one must be careful to appreciate the
  specific meaning of such terms intended by the author.

  \subsection{Some topics not covered}\label{secnotCoveredCS}

  We list a number of themes that would warrant a closer inspection
  but for which we lacked time or expertise on this occasion:
  \begin{itemize}  
  \item Generating and detecting timing and rhythm patterns in music,
    speech or gesture recognition / production \citep{Eck02, Eck02a,
      Eck06}.
  \item Methods for dynamical adapation of learning rates in
    gradient-descent training of neural networks.
  \item Timescales in connection with statistical efficiency of neural
    sampling and Markov-chain Monte Carlo sampling algorithms
    \citep{Neal93, Jaeger20}.
  \item Interactions between adaptation rates, memory duration, and
    residual approximation errors in online adaptive signal processing
    \citep{FarhangBoroujeny98}.
  \item The role of delays in (neural network based) architectures for
    motion control.
  \item Subsampling and supersampling in digital signal processing.
  \item Attention and working memory mechanisms in deep learning,
    especially for language processing \citep{Bahdanauetal14}.
     
  \end{itemize}

\section{Take-homes}\label{secAdvice}

Our meandering journey through the landscape of temporal and timescale
phenomena in natural and artificial ``cognitive'' systems has
delivered a large and speckled collection of findings. We could not
bind them together in a unifying ``story'' (we tried this in a first
write-up but had to abandon the attempt because there were many themes
left that did not fit into the unified picture that we started to
draw). But despite the heterogeneity of our findings, there are some
lessons learnt that we believe provide useful input to the MemScales
consortium at an early time in the project:

\begin{description}
\item[Timescales is a multidimensional concept.] There are many ways
  in which ``timescale'' themes come to the surface when thinking
  about cognitive systems. One consequence for hardware and
  computational methods research in MemScales: there is not a single
  good (or even best) way to design neuromorphic systems with regards
  to timescales. How multiple timescales have to be physically and
  algorithmically supported depends on the use scenario of the
  targeted system. 
\item[Timescales cannot be ignored.] Our belief that timescales and
  memory hierarchies are important was a \emph{raison d'\^{e}tre} for
  launching our project. Our findings substantiate and underline this
  initial belief and convince us that a dedicated project focus on
  timescales is a necessary topic of dedicated research in the further
  development of neuromorphic computing.
\item[More complex cognitive processing needs more timescales.] A
  task's cognitive complexity seems closely linked to the spectrum of
  memory timescales needed for it. This indicates that for a
  systematic development of neuromorphic technologies it is helpful to
  work out a complexity hierarchy of task types and initially not
  ``reach for the stars'' but concentrate on tasks of modest
  complexity that require to integrate information only across a few
  timescales only (or even a single one).
\item[Relative and absolute timescales.] A neuromorphic computing
  system (hardware plus algorithms) must support a range of timescales
  that widens as the cognitive task complexity grows. If the system is
  used in offline mode (for instance, text processing), one only needs
  to aim for a wide range of relative timescales. If the system is
  targeting online processing tasks (for instance robot control or
  cardiac monitoring), in addition one must match the system's
  absolute timescales to the task data streams. The main challenge
  here is probably to physically realize slow enough timescales. 
\item[Tricks to avoid many physical timescales.] It is not easy to
  realize a wide spectrum of timescales directly in physical effects
  on a non-digital neuromorphic microchip. There are a number of
  workarounds that may alleviate this challenge:
  \begin{itemize}
  \item Digital-analog hybrid processors where slow timescales are
    made possible by digital buffering. Needs a development of
    dedicated digital-analog algorithmics.
  \item Large (possibly \emph{very} large) RNNs can encode large
    amounts of information from past input in the current network
    state and thus have longer dynamic memory spans. Needs microchip
    technologies for realizing (very) large RNNs.
  \item Reservoir transfer methods may have some potential for
    realizing long memory spans even in modestly sized RNNs if these
    are explicitly trained for the specific memory functionalities
    demanded by the target task.
  \item Designing RNN architectures that include explicit mode
    switching mechanisms (possibly trainable) may realize temporally
    nested processing levels. Needs the development of dedicated
    architectures and learning algorithms, and a clear understanding
    of the ``stack memory'' demands of a task.
  \end{itemize}     
\item[Delays may make neuromorphic computing difficult.] Signal travel
  delays in unclocked analog neuromorphic microchips become a problem
  when delay times are not well separated from the fastest timescales
  demanded by the processing task (in which case delays can be
  ignored). For high-frequency online processing tasks (for instance
  in future neuromorphic low-energy communication nodes), an explicit
  modeling and algorithmic compensation for physical delays is needed.
  For multi-timescale offline tasks, an upper limit for task
  throughput rates is given by the necessity to separate physical
  delays from the fastest task timescale.
  
  We note that delays are no mathematical or algorithmical problem in
  digital computing as long as physical on-chip delay times are much
  shorter than clock cycle times.
  
\item[Delays may make neuromorphic computing easy.] If one would find
  a way to physically realize tapped delay lines (by traveling waves
  or solitons, maybe skyrmionic?), multiple timescale dynamics (with
  longest scale given by longest signal travel time on the delay line)
  might become explicitly designable. Needed: mathematics and
  algorithms embedding tapped delay lines in analog computing
  architectures.  
\item[Life history timescale.] If the motto of brain-like computing is
  taken seriously, the ``lifespan'' timescale of an individual
  hardware system becomes relevant. While digital microchips don't age
  and don't have an individuation history: if they start processing
  0's and 1's differently from when they were sold, they are called
  ``broken'' and are replaced by an identical twin. Analog
  neuromorphic microchips will likely be individual from the moment
  when they leave the fab (due to device mismatch); they will often
  exhibit slow parameter drift and physical aging; and they cannot be
  ``programmed'' in the traditional sense but will likely have to be
  trained. This will lead to individual lifelong learning and
  adaptation histories. Needed: novel mathematical tools to describe
  qualitative change and continual/lifelong learning schemes
  (algorithms and training schemes) that are appropriate for
  physically aging systems, which in particular will require a
  collaboration between learning and homeostatic self-stabilization
  mechanisms.
\end{description}

\section{Timescale requirements for neuromorphic hardware designs
  for STDP and RC processing} \label{secGuides}

Today, virtually all analog spiking neuromorphic hardware
demonstrations are based on either STDP, RC, or sometimes a
combination of the two. This also holds true for our research in the
predecessor project NeuRAM3 (for instance \cite{Heetal19,
  Yousefzadehetal18, Covietal18}). Although our survey has made it
clear that biological brains as well as machine learning techniques
derive their strength from a much wider range of computational /
learning principles than STDP and RC, at the current point in time it
makes sense to focus on these two paradigms, identify the current
state of development in the theory and the practical uses thereof, and
from that derive concrete (minimal) requirements for physical
timescales that have to be delivered by analog spiking neuromorphic
hardware. We treat STDP and RC in turn, but start with a general
summary of timescales and their biological, algorithmical and hardware
realizations.

\subsection{Timescales: biological, algorithmical, hardware}

Research in the neurosciences has identified a plethora of neural
adaptation mechanisms. They are based on a wide spectrum of physical
and physiological mechanisms and operate on all levels of the brain's
hierarchical architecture, from synapses to membranes to entire neural
assemblies and projection pathways; and they serve many functions (as
far as one can identify them today) in homeostatic regulation, fast
and slow adaptation to input characteristics, short-term, working and
long-term memory, learning and ontogenesis. This richness is far from
being fully understood in the neurosciences, and there exists no
unified or comprehensive mathematical model. 

Nonetheless, it is instructive to be aware of some core concepts and
findings from neuroscience. Table 1 gives a highlevel indicative
overview which reflects the ongoing discussions in
the consortium. It remains to be explored which physical effects of
hardware devices can serve which biologically motivated
mechanism. This is an intricate question because it is not the
physical device / effect per se that serves a computational/biological
mechanism, but a complex interplay of the core physical effect with
circuit designs and control schemes, like for instance pulse pattern
schemes for setting PCM resistances.

\vspace{0.5cm}
\begin{center}
\begin{tabular}{|p{3cm}|p{3cm}|p{2.5cm}|p{4.5cm}|}
  \hline
  \textbf{Biological plasticity phenomenon} & \textbf{Timescale} &
                                                                  \textbf{Mechanism}
  & \textbf{Candidate physical device / effect} \\
  \hline  \hline
  Short-term plasticity  & 1 ms -- 10 ms & STDP, SDSP & capacitors \\ \hline
  Long-term plasticity & 10 ms -- 500 ms for weight change; 1 h --
                         years for weight preservation  & LTP/LTD & non-volatile
                                                   memristive devices
                                                   (for preserving the
                                                   results of LTP) \\ \hline
  Intrinsic plasticity & 0.5 s -- 10 s & threshold adaptation  & volatile
                                                           ReRAM, TFT, ...\\ \hline
                          
Homeostatic plasticity  & 1 s -- 1 h & synaptic scaling & volatile
                                                         ReRAM, PCM drift, TFT, ...\\ \hline
  
Structural plasticity & 1 h -- lifetime & architecture reorganisation &
                                                                      reconfigurable
/ extendable  architectures \\ \hline
\end{tabular}
\end{center}

\begin{center}{\textbf{Table 1} Overview of plasticity
  phenomena}\end{center}

The large number of physiological mechanisms underlying this spectrum
of phenomena, as well as the wealth of formal models in theoretical
neuroscience that capture these phenomena at different levels of
abstraction, as well as physical differences between brain physiology
and electronics, make it impossible to copy biological mechanisms 1-1
to electronic microchips. Furthermore, it is not necessarily the 
most promising engineering strategy to even try to copy brain
mechanisms exactly into analog spiking hardware. On the one hand, many
biochemical mechanisms will be hard to replicate in electronic
systems, and on the other hand, electronic systems may offer
opportunities (especially, faster timescales or very long-term
non-volatile memory states) that are not affored in physiological
brain substrates. Yet, Table 1 teaches a clear lesson: in order to
endow artificial with proxies of the biological inference, adaptation
and learning mechanisms, a wide range of timescales must be covered.

How this is concretely done will depend on the available hardware,
targeted performance and use-cases, and algorithmic models. In the
following subsection we will work this out in an examplary case study.    

\subsection{Timescale requirements resulting from demands for STDP
  learning: a worked-out case study} \label{secGuidesSTDP}

There is not a single, well-defined STDP adaptation rule in biological
brains. In fact, it is an experimental challenge to localize, measure
and formulate STDP mechanisms in mammalian brains.  In the machine
learning / computational neuroscience / neuromorphic engineering
communities, a broad variety of STDP variants and combinations of them
with other neural adaptation mechanism have been explored ---
\cite{JoshiTriesch09, clopath2010connectivity,GraupnerBrunel12,
  GaltierWainrib13, KlampflMaass13, Roclinetal13, Bengioetal17,
  Thieleetal18} are but a small selection of approaches that document
this variability. The initial specific concept of STDP (as described
in the landmark paper by \cite{Markrametal97}, with many forerunners)
does not cover this variability. The term ``spike-timing-dependent
synaptic plasticity'' and the acronym STDP was introduced in
\cite{Songetal00}. The term \emph{Spike-Driven Synaptic Plasticity}
(SDSP), apparently introduced by \cite{Fusietal00} in a formal model
of an adaptive synapse independent of, but potentially effective in a
variety of learning/adaptation mechanisms, should be preferred over
the term STDP when one considers spike-driven synaptic plasticity
phenomena in a more general setting than the original STDP
framing. Since neuromorphic electronic circuits and neural network learning
algorithms used in them explore and exploit more general mechansism
than STDP proper, we will use the term SDSP
in this section.

In order to become concrete, we
must however settle on a specific model, and this should not be a
repetition of what we already developed in NeuRAM3. Instead, our
choice should open doors for the currently most promising line of SDSP
exploits, the so-called 3-factor rules. Again, this principle
comprises many different variants. Generally speaking, in 3-factor
SDSP rules, the adaptation effects determined by the basic two factors
(pre- and postsynaptic activations) of SDSP become multiplicatively
modulated by a third factor, which represents some kind of global
control signal, which can be variously interpreted as a reward signal,
a derivate of a supervised target signal, a temporal coordination /
synchronization guide, or a mean-field population activity signal for
achieving homeostatic regulation of a neuron's average activity level
(summarized in \cite{Kusmierzetal17}).

Our choice is to opt for the first among the mentioned
interpretations, and concretely for the model recently proposed by
\cite{Bellecetal19, Bellecetal20}. This model, named the \emph{e-prop}
model, imports mechanisms from reinforcement learning and utilizes
them to realize an approximation of stochastic gradient descent (SGD)
with an SDSP mechanism. SGD is the main enabling learning principle
that empowers deep learning techniques, and is thus of great potential
value for neuromorphic technologies since the current state of the art
in machine learning is defined through SGD trained neural
networks. However, in the deep learning field, one does not use
spiking neuron models. Much effort has been spent in the last 10 years
to find approximations of SGD that also work in spiking neural
environments, with limited success. The model of Bellec et al.\ has
immediately created a strong resonance, building on and transcending
previous approaches to apprximate SGD in spiking networks, is
mathematically transparent, can be adopted to a variety neuron models,
has been explicitly formulated with analog spiking hardware
implementations in mind, and furthermore MemScales members (Indiveri,
Jaeger) enjoy a long-standing collaboration with the group of Wolfgang
Maass where this model originates. Closely related SDSP-realized
approximations of SGD are currently being explored in a number of
research groups. \cite{NeftciAverbeck19} review approaches of
transferring neurobiological models of reinforcement learning to
artificial neural networks, emphasizing the benefits of neuron models
that include sub-mechanisms that operate on different timescales, and
report brackets for biological time constants. \cite{Payvandetal20}
(whose first author is a member of the INI) present an analog circuit
for an on-chip realization of (a version of) such learning rules, and
demonstrate it in a simulation. Concrete values of effective time
constants are unfortunately not provided. The documentation of
mathematical formalism in \cite{Bellecetal19, Bellecetal20} is
particularly detailed, which gives us the option to analyse conditions
on time constants, which we now proceed to do.

Following \cite{Bellecetal20}, we first give a brief summary of e-prop
for the case of leaky integrate-and-fire (LIF) neurons (formulated in
a discrete-time setting, using a unit timestep of $\delta t = 1$
millisecond), the most simple and arguably most popular spiking neuron
model in neuromorphic engineering theory. The core of SGD algorithms
in supervised learning for the adapation of a synaptic weight $w_{ji}$
from pre-synaptic neuron $i$ to post-synaptic neuron $j$ is the error
gradient $\frac{dE}{dw_{ji}}$, which can be factorized as

\begin{equation}\label{eBellec1}
  \frac{dE}{dw_{ji}} = \sum_{t} \frac{dE}{dz^t_j} \cdot
  \left[\frac{dz^t_j}{dw_{ji}}\right] =: \sum_{t}  L_j^t\,  e^t_{ji},
\end{equation}
where $z^t_j$ is the postsynaptic spike train (a binary signal), the
summation goes over the time points of the learning history, the 
factor ${dE}/{dz^t_j} =: L_j^t$ is the \emph{learning signal}, and the
factor ${dz^t_j}/{dw_{ji}} =: e^t_{ji}$ is the
\emph{eligibility trace}. The elegibility trace depends on pre- and
postsynaptic spiking (see below) and are thus  a form of SDSP. 
The learning signal is the ``third'' factor
in the customary terminology when one speaks of 3-factor rules. 

Note that this formulation \eqref{eBellec1} 
captures the weight change gradient obtained from accumulating
information about a whole training sequence or a training batch. For
an instantaneous weight adapation in a single model update step from
time $t$ to $t + \delta t = t + 1$, as needed for adaptive hardware
implementations, \eqref{eBellec1}  reduces to the
online learning rule
\begin{equation}\label{eBellec1a}
  \Delta^t w_{ji} = - \eta \; L_j^t\,  e^t_{ji},
\end{equation}
where $\eta$ is a learning rate. We now take a closer look first at
the elibility trace and the ``third factor'', the learning signal, in
that order. We first give an brief summary account of the formalism in
\cite{Bellecetal20}, which is geared toward discrete-time
simulations on a digital computer, and then discuss what conditions on
physical time constants in unclocked event-based analog hardware
implementations can be derived.

Since spike pre- or postsynaptic spike
trains $z^t_j$ are not differentiable, they are  replaced by 
exponentially smoothed filtered versions
\begin{equation}\label{eBellec2}
  \bar{z}^t := \mathcal{F}_\alpha (z^t) := \alpha \,
  \mathcal{F}_\alpha (z^{t-1}) + z^t
\end{equation}
when needed. \cite{Bellecetal20} derive that the eligibility trace
$e^t_{ji}$ can then be re-written as
\begin{equation}\label{eBellec3}
  e^{t+1}_{ji} = \psi_j^{t+1}\; \bar{z}^t_i, 
\end{equation}
where $\psi_j^{t}$ is a pseudo-derivative of $\partial z^t_j
\, / \, \partial v_j^t$ (used variously in the
literature for making spike trains differentiable under consideration
of the post-synaptic neuron's $j$ refractory period $r$;
\cite{Bellecetal20} refer back to \cite{Bellecetal18}), given by
\begin{equation}\label{eBellec4}
  \psi_j^{t} := \left\{ \begin{array}{cl}
                        0 & \mbox{for } t \mbox{ inside }r\\
                        \frac{1}{v_{\mbox{\scriptsize
                        th}}}\;\gamma_{\mbox{\scriptsize pd}}\,\max
                        \left(0, 1 - \left| \frac{v_j^t -
                          v_{\mbox{\scriptsize
                          th}}}{v_{\mbox{\scriptsize th}}}\right|
                          \right)  & \mbox{else,}
                      \end{array}
                    \right.
\end{equation}
where in turn $v_j^t$ is the membrane potential of neuron
$j$, $v_{\mbox{\scriptsize th}}$ its firing threshold, and
$\gamma_{\mbox{\scriptsize pd}}$ is a heuristic damping parameter that
is set to $\gamma_{\mbox{\scriptsize pd}} = 0.3$ by Bellec et al; the role of
this damping parameter is to improve numerical stability of
approximated gradient descent in networks that have many layers. The
membrane potential $v_j^t$ evolves according to
\begin{equation}\label{eBellec5}
  v_j^{t+1} = \alpha v_j^{t} + \sum_{i \neq j} W^{\mbox{\scriptsize
      rec}}_{ji}\, z_i^t + \sum_i W^{\mbox{\scriptsize
      in}}_{ji} \; x_i^{t+1} - z_j^t \, v_{\mbox{\scriptsize th}},
\end{equation}
\begin{equation}\label{eBellec6}
  z_j^t = H(v_j^t - v_{\mbox{\scriptsize th}}),
\end{equation}
where $W^{\mbox{\scriptsize rec}}_{ji}, W^{\mbox{\scriptsize
    in}}_{ji}$ are the recurrent and input weights to neuron $j$, $x^t$
is the input signal and
$H$ is the Heaviside step function. The decay rate $\alpha$ is can be
expressed in terms of an exponential decay function by
\begin{equation}\label{eBellec7}
 \alpha = \exp(-\delta t / \tau_m),
\end{equation}
where $\tau_m$ is the membrane time constant. A biologically plausible
value is $\tau_m = 20$ ms. With a stepsize $\delta t = 1$ ms (which is
used in the simulations in \cite{Bellecetal20}), this gives $\alpha
\approx 0.95$.

The learning signal $L_j^t$ in \eqref{eBellec1} measures the deviation
between the output signals $y_k$ generated by the network output neurons $k$ (which are not recurrently connected
to each other), defined by the leaky integration rule
\begin{equation}\label{eBellec8}
y_k^t = \kappa y_k^{t-1} + \sum_j  W^{\mbox{\scriptsize out}}_{kj}\;
z^t_j + b^{\mbox{\scriptsize out}}_k, 
\end{equation}
and the target outputs $y^{\ast, t}_k$ by a simple linear combination
of the errors
\begin{equation}\label{eBellec9}
L_j^t = \sum_k B_{jk}(y_k^t - y^{\ast, t}_k).
\end{equation}
The error backprojection weights $B_{jk}$ are determined in
\cite{Bellecetal20} according to various heuristics, among them fixing
them at random values. While this works satisfactorily in the
demonstrations given in \cite{Bellecetal20}, we see this dependence on
heuristic intuition to define the learning signal as an opportunity
for further improvements of this model. Which values of $\kappa$ were
chosen in the demonstrations in \cite{Bellecetal20} remained
un-documented there. However, it is clear that for tasks where the
target outputs $y^{\ast, t}_k$ are smooth signals, they must be
assumed to be high-pass filtered to preclude arbitrary baseline drifts
which cannot be learnt by neuronal outputs. To connect our following
discussion of synaptic/neuronal time constants with task-specific time
constants, we will consider the period length $T^\ast$ (in
milliseconds) of the lowest significant frequency in $y^{\ast, t_k}$
as the slowest task-relevant timescale. The challenge for online
learning with spiking neurons is to be slow enough to be able to
integrate task-relevant information on that timescale $T^\ast$.

We now consider the question how this model, which is formulated in a
discrete-time set-up for simulation on digital computers, translates
into requirements for RNN implementations in unclocked, event-based,
spiking hardware.

We first consider the second factor $\bar{z}^t_i$ in the eligibility
trace \eqref{eBellec3}, which represents pre-synaptic spikes arriving
at the synapse $w_{ji}$. Note that the $1$ ms time difference between $t+1$
and $t$ in \eqref{eBellec3} is due to the discrete-time simulation
scenario, where a unit time step is assumed for propagating the
information from neuron $i$ to neuron $j$. For electronic event-based
neuromorphic hardware we may assume that the travel delays of electric
signals are negligible, hence instead of \eqref{eBellec3} we will
consider
\begin{equation}\label{eBellec10}
  e^{t}_{ji} = \psi_j^{t}\; \bar{z}^t_i. 
\end{equation}

According to \eqref{eBellec2} and \eqref{eBellec7}, $\bar{z}^t_i$ is
an exponentially smoothed version of $z_i^t$ with an exponential time
constant that we will call $\tau_{\mbox{\scriptsize \sf pre}}$. In
order to exploit \eqref{eBellec2} in analog unclocked hardware, a
physical variable available at the physical implementation of synapse
$w_{ji}$ must represent $\bar{z}^t_i$, that is, a physical leaky
integration of the incoming spike train ${z}^t_i$ with time constant
$\tau_{\mbox{\scriptsize \sf pre}}$ must be effected somewhere in the
circuit --- either at the sending neuron $i$ (then this signal must be
sent to all receiving neurons $j$), or at the receiving synapse
$w_{ji}$ (then the integration must by physically repeated at all
synapses to which $i$ sends out its spike train).

We note that it is not possible to derive general rules for how
$\tau_{\mbox{\scriptsize \sf pre}}$ should be set for the network to
solve its learning task. Whether a specific setting will be successful
depends on many design variables, for instance the size of the RNN (in
larger RNNs, less precision per synapse is needed), functional
specialization of neurons $i$ and $j$ (they might specialize on
high-frequency components in the outward task, leading to more relaxed
constraints on $\tau_{\mbox{\scriptsize \sf pre}}$), the average
firing rate of the feeding neuron $i$ (the higher, the smaller can
$\tau_{\mbox{\scriptsize \sf pre}}$ be), and importantly, a model of
how task-relevant information is encoded in  ${z}^t_i$. We must make
further assumptions to arrive at a well-defined problem. We will
proceed under the following assumptions.
\begin{enumerate}
\item Task-relevant information is encoded in the network by rate
  coding.
\item The synapse $w_{ij}$ contributes significantly to the network's
  functionality with regards to the slowest task-relevant timescale
  $T^\ast$.
\end{enumerate}
The leaky integration of ${z}^t_i$ should be such that a significant
memory trace of spikes that lie $T^\ast$ in the past is still present
in $\bar{z}^t_i$. What ``significant'' means is subject to an
essentially arbitrary commitment. Here we opt for a plausible
heuristic and require that the contribution of 
${z}_i^{t-T^\ast}$ to $\bar{z}^t_i$ is reduced by a \emph{forgetting
  factor} $F$ of at most $1/2$ at
time $t$. This leads to the condition
\begin{equation}\label{eBellec11}
  \exp(- T^\ast / \tau_{\mbox{\scriptsize \sf pre}} ) \geq 1/2,
\end{equation}
that is
\begin{equation}\label{eBellec12}
  \tau_{\mbox{\scriptsize \sf pre}} \geq T^\ast \, \frac{-1}{\log(1/2)}
  \approx 1.4 \cdot T^\ast.
\end{equation}

Next we turn to the first factor $\psi^t_j$ in the eligibility trace
\eqref{eBellec10}. This factor accounts for the postsynaptic spike
timing when interpreted in an SDSP perspective. Inspecting
\eqref{eBellec4} we see that this factor follows the temporal profile
of the membrane potential $v^t_j$ of neuron $j$, which in turn (see
\eqref{eBellec5}) is a leaky integration of recurrent and input spike
trains arriving at $j$.  We have to transfer the discrete-time
formulation of Bellec et al.\ to the continuous-time, event-based
situation in analog unclocked hardware, that is, we must translate the
discrete timestep discount factor $\alpha$ in \eqref{eBellec5} to an
exponential decay rate that we will call (like Bellec et al.\ do)
$\tau_m$. Again we must make additional assumptions to arrive at a
specific statement of our problem. Repeating the assumptions and the
heuristic that we committed above, we arrive at the same conclusion as
in \eqref{eBellec12}:
\begin{equation}\label{eBellec13}
  \tau_m \geq T^\ast \, \frac{-1}{\log(1/2)}
  \approx 1.4 \cdot T^\ast.
\end{equation}

This suggests that membrane leaking time constants are needed that are
in the order of the slowest relevant task time constants.
\cite{Bellecetal20} used a biologically motivated value of
$\tau_m = 20$ ms. They demonstrated their model on a supervised task of
phoneme recognition, where $T^\ast$ is $10$ ms, which satisfies our
constraints \eqref{eBellec12} and \eqref{eBellec13}. However, in
another experiment, where e-prop was adapted to a reinforcement
learning situation, the task-relevant slowest timescale was in the
order of $T^\ast = 2000$ ms, still with $\tau_m = 20$ ms. This is at
odds with \eqref{eBellec13}. The solution to this puzzle is an
argument that combines the influence of network size with the choice
of the forgetting factor $F = 1/2$. If we plug in smaller forgetting
factors in \eqref{eBellec12}, \eqref{eBellec13}, we end up with
smaller admissible time constants
$\tau_{\mbox{\scriptsize \sf pre}}, \tau_m$. They result in smaller
$T^\ast$-delayed task-relevant additive components in the signals
$ \bar{z}^t_i, \psi_j^{t}$, a source of variation which in turn can
however be compensated by the linear combinations effected through
$W^{\mbox{\scriptsize rec}}_{ji}, W^{\mbox{\scriptsize
    in}}_{ji}$. This efficacy of this compensation scales with the
size $N$ of the RNN and the numerical accuracy of the used computing
environment. Bellec et al.\ used floating-point precision arithmetics
and large networks (with 2400 neurons in the phoneme recognition demo,
not documented for the reinforcement learning task). In this light, our
suggestions \eqref{eBellec12}, \eqref{eBellec13} should be considered
as extremely conservative if not pessimistic, relevant (only) for very
small networks with a few neurons and low numerical precision (or with 
noise).  

We summarize our findings:

\begin{itemize}
\item For implementing the e-prop algorithm for supervised training or
  RNNs in analog spiking neuromorphic hardware on the basis of
  elementary LIF neuron models, two leaky integration mechanisms are
  needed, one for the membrane potential and one for the 
  smoothing of spike trains arriving at a synapse.
\item Lower bounds on the minimally necessary time constants for these
  two integration mechanisms depend on a number of design variables
  (in particular network size and realizable numerical accuracy) and
  task specifics (in particular slowest task-relevant time constant in
  task signals). Under the most conservative assumptions (very small
  network, low numerical accuracy) one can reason that the leaky
  integration time constants for the membrane and synapse integrations
  must be in the order of the slowest task-specific time constant. As
  network size and/or available numerical accuracy increases,
  increasingly faster neuronal/synaptic time constants can be expected
  to be sufficient for realizing the e-prop algorithm.
\item Not all membrane or synapse time constants need to satisfy the
  conditions described here. In order to enable a RNN architecture to
  cope with the slowest task-relevant timescales, it is enough if
  \emph{some} neurons / synapses are capable of the required slow
  integrations. Specifically, hierarchical network architectures
  are often designed in a way that ``higher'' layers operate in slower
  timescale modes than ``lower'' layers.  
\end{itemize}

We emphasize that the considerations made above are tied to the
specifics of the e-prop algorithm, with its specific version of SDSP
and its specific training objective and system architecture proposed
in \cite{Bellecetal20}. There are
many other SDSP rules, other training objectives (in particular,
unsupervised ones, or tasks based on non-temporal data) and
architectures, where other considerations would have to be done. In
particular, the necessity of leaky-integrating incoming spike trains
at each synapse is a consequence of the specific e-prop mathematics
and will not be required in many other SDSP versions, tasks or
architectures.

The main lesson to be drawn from this case study is that there should
be a mechanism in the neuromorphic system whose time constant matches
the slowest timescale $T^\ast$ of the outward task (where ``matching''
needs to be qualified, it need not be identity but can mean that the
corresponding hardware mechanism has a faster timescale that can be
expanded to the task timescale through computational effects). We
found the same lesson taught to us in a quite different experimental
and algorithmic scenario too, as will be reported in Section
\ref{secTimeScaleRC}. If that lesson holds true, then a very wide
range of task-dictated timescales $T^\ast$ must be served: ranging
from milliseconds in robot/prosthetics control to days or weeks or even
years in environmental monitoring, just to name two application tasks
that have been proposed as targets for neuromorphic computing
technologies. 

We emphasize that this case study does not imply a recommendation to
for MemScales research to implement this specific model. We chose it
as a representative because the article of \cite{Bellecetal20} gave a
mathematical model in all detail, from which we could develop an
exemplary analysis. Other SDSP models have been or are being explored
in our consortium, like \cite{Yousefzadehetal18} or
\cite{Cartigliaetal20}. It is impossible to provide a theoretical
analysis that covers all options, and it would be inappropriate
to try to identify a ``best'' one.  

We finally point out that the e-prop algorithm has recently been
employed in a reservoir computing set-up, where it was used to
optimize the recurrent weights of a reservoir for an entire class of
learning tasks in a ``learning to learn'' scenario
\citep{Subramoneyetal21}.

\subsection{Timescale requirements for RC systems based on analog
  spiking event-based neuromorphic hardware} \label{secTimeScaleRC}

Reservoir computing based on physical reservoirs is a flourishing
research area. Physical RC systems have been built on the basis of
many different non-digital physical substrates. Popular media include
optics \citep{Antoniketal18}, nano-mechanics \citep{Coulombeetal17},
carbon nanotubes \citep{Daleetal16}, magnetic skyrmions
\citep{Prychynenkoetal18}, spintronics \citep{Torrejonetal17}, or gold
nanoparticle thin films \citep{Minnaietal18} (survey in
\cite{Tanakaetal18}). These studies are mostly
experimental. Theoretical analyses, or at least systematic
explorations of dynamical phenomena in controlled simulations, are
scarce. We are aware only of one work in the optical RC community
\citep{Grigoryevaetal16} and another one in memristive electronics
based reservoirs \citep{Sheldonetal20}, which is however still rather
rudimentary.

An inherent obstacle to general theoretical analyses is
that every physical systems comes with ideosyncratic dynamical
properties that leave their mark on the computational properties of
the respective system, and would have to be analysed on a case by case
basis. While a large body of analytical research has accumulated over
the last two decades for reservoirs that are mathematically defined on
the simplest possible rate-based neuron model (the echo state
networks), insights made there do not easily carry over to other sorts
of reservoirs. Specifically, no theoretical analyses of computational
/ learning characteristics of reservoirs based on analog spiking
continuous-time neural networks are yet available.

A natural starting point for such analyses, with special attention
paid to timescale phenomena, would be to study the \emph{memory
  capacity} (MC) of analog spiking reservoir RNNs. In its original
format, which was expressed for discrete-time non-spiking reservoirs
of the \emph{echo state network} type, the MC is a measure for how
many previous inputs of a one-dimensional white noise signal can be
recovered by trained linear readouts, weighted with an accuracy
factor. In the work that started this research line \citep{Jaeger02b}
it was shown that MC is bounded by the number of neurons in the
reservoir. This triggered a large number of follow-up studies (a
Google Scholar query on {\sf ``echo state network'' ``memory
  capacity''} returns more than 600 papers) which extended the
original analysis with regards to input signal type, neuron model,
noise robustness, input dimension, network architecture, alternative
definitions of MC, and more. Contributions came from mathematics,
theoretical physics, the neurosciences and machine learning. The broad
interest in this question can be explained by the fundamental nature
of the question of information transport in dynamical systems in
general, the relevance for machine learning tasks (see
\cite{Dambreetal12} for the intimate connection between memory
properties and general computational capacities), and the relevance
for understanding dynamical short-term memory in biological brains. We
are however not aware of mathematical analyses of MC in
spiking RNNs, although it has been experimentally measured in a number
of simulation studies.

In the group of Jaeger at the University of Groningen, the PhD student
Dirk Doorakkers, whose position is funded through MemScales and who is
a mathematician with a specialization in dynamical systems theory (and
who authored Section \ref{secMath} of this deliverable report), will
carry out a dissertation project that centrally addresses the question
of information transport in spiking RNNs. His project, with the
working title \emph{Double transients in multi-timescale systems
  provide a geometric description for dynamic coding with
  activity-silent working memory}, plans a rigorous analysis of
mechanisms in spiking RNNs where
\begin{enumerate}
\item an input signal is initially \emph{encoded} in a temporal
  activation pattern of the RNN, which
\item propagates in time through the RNN for a delay (``memory'')
  period $d$, undergoing a sequence of transformations, until
\item upon a cue signal a desired output transform of the input signal
  is recovered by a \emph{decoding} (``readout'') mechanism.       
\end{enumerate}
This analysis will be done with the tools of contemporary dynamical
systems theory, in particular slow-fast systems (singular perturbation
methods) and bifurcation theory, aiming for a characterization of such
memory mechanisms in terms of generic geometrical dynamical  systems
concepts, which to a certain degree would render the analyses
transferable to general classes of multi-timescale hardware
reservoirs. This will constitute a substantial contribution to task T
1.4, \emph{Toward a general model of unconventional computing}.

For the time being, the best that we can offer is a summary of
findings that we collected in the NeuRAM3 forerunner project to
MemScales. Jaeger's group was charged to realize an online heartbeat
anomaly classifier on the Dynap-se, a spiking analog
neuromorphic microchip developed at the Institute of Neuroinformatics
in Zurich \citep{Moradietal18}. The challenge was that the natural
time constant of human heartbeats is 1 sec, while the slowest
available time constants for spike train integration on the Dynap-se
were much faster. Our findings and methods are reported in
\cite{Heetal19}. Here we give a summary account, which agrees well
with our observations in Section \ref{secGuidesSTDP}:
\begin{itemize}
\item In earlier simulations (not reported in \cite{Heetal19}) we
  found that the learning task was possible with spike train
  integration time constants that matched the task time constants.
\item The unavailability of physical time constants that were as slow
  as the 1 sec time constant of the heartbeat data led to failures in
  ``direct-attack'' attempts to train a Dynap-se based reservoir.
\item The task became solvable on the Dynap-se when a novel
  \emph{reservoir transfer} method was employed to pre-configure the
  synaptic weights in the hardware reservoir in a way that allowed
  linear combinations of spike trains arriving at a receiving neuron
  to compensate for the small forgetting factors (see Section
  \ref{secGuidesSTDP}) inherent in the Dynap-se physics. The reservoir
  comprised about 750 neurons.
\end{itemize}


\clearpage
\begin{thebibliography}{158}
\providecommand{\natexlab}[1]{#1}
\providecommand{\url}[1]{\texttt{#1}}
\expandafter\ifx\csname urlstyle\endcsname\relax
  \providecommand{\doi}[1]{doi: #1}\else
  \providecommand{\doi}{doi: \begingroup \urlstyle{rm}\Url}\fi

\bibitem[Albus(1993)]{Albus93}
J.~S. Albus.
\newblock A reference model architecture for intelligent systems design.
\newblock In P.~J. Antsaklis and K.~M. Passino, editors, \emph{An Introduction
  to Intelligent and Autonomous Control}, chapter~2, pages 27--56. Kluwer
  Academic Publishers, 1993.

\bibitem[Allen(1991)]{Allen91}
J.~F. Allen.
\newblock Time and time again: The many ways to represent time.
\newblock \emph{International Journal of Intelligent Systems}, 6\penalty0
  (4):\penalty0 341--355, 1991.

\bibitem[Antonik et~al.(2018)Antonik, Hermans, Haelterman, and
  Massar]{Antoniketal18}
P.~Antonik, M.~Hermans, M.~Haelterman, and S.~Massar.
\newblock Random pattern and frequency generation using a photonic reservoir
  computer with output feedback.
\newblock \emph{Neural Processing Letters}, 47\penalty0 (3):\penalty0
  1041--1054, 2018.

\bibitem[Baddeley(2003)]{Baddeley03}
A.~Baddeley.
\newblock Working memory: looking back and looking forward.
\newblock \emph{Nature Reviews: Neuroscience}, 4\penalty0 (10):\penalty0
  829--839, 2003.

\bibitem[Bahdanau et~al.(2015)Bahdanau, Cho, and Bengio]{Bahdanauetal14}
D.~Bahdanau, K.~Cho, and Y.~Bengio.
\newblock Neural machine translation by jointly learning to align and
  translate.
\newblock In \emph{International Conference on Learning Representations
  (ICLR)}, 2015.
\newblock URL \url{http://arxiv.org/abs/1409.0473v6}.

\bibitem[Bargmann(2006)]{bargmann2006chemosensation}
Cornelia~I Bargmann.
\newblock Chemosensation in c. elegans.
\newblock In \emph{WormBook: The Online Review of C. elegans Biology
  [Internet]}. WormBook, 2006.

\bibitem[Barkai and Leibler(1997)]{barkai1997robustness}
Naama Barkai and Stan Leibler.
\newblock Robustness in simple biochemical networks.
\newblock \emph{Nature}, 387\penalty0 (6636):\penalty0 913--917, 1997.

\bibitem[Bedny et~al.(2015)Bedny, Richardson, and Saxe]{bedny2015visual}
Marina Bedny, Hilary Richardson, and Rebecca Saxe.
\newblock “visual” cortex responds to spoken language in blind children.
\newblock \emph{Journal of Neuroscience}, 35\penalty0 (33):\penalty0
  11674--11681, 2015.

\bibitem[Bellec et~al.(2018)Bellec, Salaj, Subramoney, Legenstein, and
  Maass]{Bellecetal18}
G.~Bellec, D.~Salaj, A.~Subramoney, R.~Legenstein, and W.~Maass.
\newblock Long short-term memory and learning-to-learn in networks of spiking
  neurons.
\newblock arxiv manuscript, 2018.
\newblock URL \url{https://arxiv.org/abs/2006.12484}.

\bibitem[Bellec et~al.(2019)Bellec, Scherr, Hajek, Salaj, Legenstein, and
  Maass]{Bellecetal19}
G.~Bellec, F.~Scherr, E.~Hajek, D.~Salaj, R.~Legenstein, and W.~Maass.
\newblock Biologically inspired alternatives to backpropagation through time
  for learning in recurrent neural nets.
\newblock arxiv manuscript, 2019.
\newblock URL \url{https://arxiv.org/abs/1901.09049}.

\bibitem[Bellec et~al.(2020)Bellec, Scherr, Subramoney, Hajek, Salaj,
  Legenstein, and Maass]{Bellecetal20}
G.~Bellec, F.~Scherr, A.~Subramoney, E.~Hajek, D.~Salaj, R.~Legenstein, and
  W.~Maass.
\newblock A solution to the learning dilemma for recurrent networks of spiking
  neurons.
\newblock \emph{Nature Communications}, 11\penalty0 (1):\penalty0 1--15, 2020.

\bibitem[Bengio et~al.(2017)Bengio, Fischer, and Wu]{Bengioetal17}
T.~Bengio, Y. abd~Mesnard, S.~Fischer, A. abd~Zhang, and Y.~Wu.
\newblock Stdp-compatible approximation of backpropagation in an energy-based
  model.
\newblock \emph{Neural Computation}, 29\penalty0 (3):\penalty0 555--577, 2017.

\bibitem[Berkes and Wiskott(2003)]{BerkesWiskott03}
P.~Berkes and L.~Wiskott.
\newblock Slow feature analysis yields a rich repertoire of complex cell
  properties.
\newblock \emph{Cognitive Sciences EPrint Archives (CogPrints)}, 2804, 2003.
\newblock URL \url{http://cogprints.org/2804/}.

\bibitem[Bernacchia et~al.(2011)Bernacchia, Seo, Lee, and
  Wang]{Bernacchiaetal11}
A.~Bernacchia, H.~Seo, D.~Lee, and X.-J. Wang.
\newblock A reservoir of time constants for memory traces in cortical neurons.
\newblock \emph{Nature Neuroscience}, 14\penalty0 (3):\penalty0 366--372, 2011.

\bibitem[Bienenstock et~al.(1982)Bienenstock, Cooper, and
  Munro]{bienenstock1982theory}
Elie~L Bienenstock, Leon~N Cooper, and Paul~W Munro.
\newblock Theory for the development of neuron selectivity: orientation
  specificity and binocular interaction in visual cortex.
\newblock \emph{Journal of Neuroscience}, 2\penalty0 (1):\penalty0 32--48,
  1982.

\bibitem[Botvinic and Plaut(2006)]{BotvinickPlaut06}
M.~M. Botvinic and D.~C. Plaut.
\newblock Short-term memory for serial order: A recurrent neural network model.
\newblock \emph{Psychological Review}, 113\penalty0 (2):\penalty0 201--233,
  2006.

\bibitem[Boyd(2010)]{boyd2010cerebellar}
CAR Boyd.
\newblock Cerebellar agenesis revisited.
\newblock \emph{Brain}, 133\penalty0 (3):\penalty0 941--944, 2010.

\bibitem[Bray(1995)]{bray1995protein}
Dennis Bray.
\newblock Protein molecules as computational elements in living cells.
\newblock \emph{Nature}, 376\penalty0 (6538):\penalty0 307--312, 1995.

\bibitem[Brooks(1989)]{Brooks89}
R.A. Brooks.
\newblock The whole iguana.
\newblock In M.~Brady, editor, \emph{Robotics Science}, pages 432--456. MIT
  Press, Cambridge, Mass., 1989.

\bibitem[B{\"u}sing et~al.(2010)B{\"u}sing, Schrauwen, and
  Legenstein]{Buesingetal10}
L.~B{\"u}sing, B.~Schrauwen, and R.~Legenstein.
\newblock Connectivity, dynamics, and memory in reservoir computing with binary
  and analog neurons.
\newblock \emph{Neural Computation}, 22\penalty0 (5):\penalty0 1272--1311,
  2010.

\bibitem[Callender(2011)]{Callender11}
C.~Callender.
\newblock Introduction.
\newblock In C.~Callender, editor, \emph{The Oxford Handbook of Philosophy of
  Time}. Oxford University Press, 2011.

\bibitem[Caporale and Dan(2008)]{caporale2008spike}
Natalia Caporale and Yang Dan.
\newblock Spike timing--dependent plasticity: a hebbian learning rule.
\newblock \emph{Annu. Rev. Neurosci.}, 31:\penalty0 25--46, 2008.

\bibitem[Cartiglia et~al.(2020)Cartiglia, Haessig, and
  Indiveri]{Cartigliaetal20}
M.~Cartiglia, G.~Haessig, and G.~Indiveri.
\newblock An error-propagation spiking neural network compatible with
  neuromorphic processors.
\newblock In \emph{Proc. 2020 IEEE International Conference on Artificial
  Intelligence Circuits and Systems (AICAS)}, pages 84--88, 2020.

\bibitem[Chicca et~al.(2014)Chicca, Stefanini, Bartolozzi, and
  Indiveri]{Chiccaetal14}
E.~Chicca, F.~Stefanini, C.~Bartolozzi, and G.~Indiveri.
\newblock Neuromorphic electronic circuits for building autonomous cognitive
  systems.
\newblock \emph{Proc. of the IEEE}, 102\penalty0 (9):\penalty0 1367--1388,
  2014.

\bibitem[Chow and Mallet-Paret(1983)]{ChowMP1983}
S.-N. Chow and J.~Mallet-Paret.
\newblock Singularly perturbed delay differential equations.
\newblock In J.~Chandra and A.C. Scott, editors, \emph{Coupled Nonlinear
  Oscillators}, pages 7--12. North-Holland Publishing Company, 1983.

\bibitem[Clopath et~al.(2010)Clopath, B{\"u}sing, Vasilaki, and
  Gerstner]{clopath2010connectivity}
Claudia Clopath, Lars B{\"u}sing, Eleni Vasilaki, and Wulfram Gerstner.
\newblock Connectivity reflects coding: a model of voltage-based stdp with
  homeostasis.
\newblock \emph{Nature neuroscience}, 13\penalty0 (3):\penalty0 344, 2010.

\bibitem[Cook(2000)]{Cook00}
S.~Cook.
\newblock The {P} versus {NP} problem.
\newblock Official problem description of the fourth millenium problem, Clay
  Mathematics Institute, 2000.
\newblock http://www.claymath.org/sites/default/files/pvsnp.pdf.

\bibitem[Cooper and Bear(2012)]{cooper2012bcm}
Leon~N Cooper and Mark~F Bear.
\newblock The bcm theory of synapse modification at 30: interaction of theory
  with experiment.
\newblock \emph{Nature Reviews Neuroscience}, 13\penalty0 (11):\penalty0
  798--810, 2012.

\bibitem[Coulombe et~al.(2017)Coulombe, York, and Sylvestre]{Coulombeetal17}
J.~C. Coulombe, M.~C.~A. York, and J.~Sylvestre.
\newblock Computing with networks of nonlinear mechanical oscillators.
\newblock \emph{PLOS ONE}, 12\penalty0 (6), 2017.
\newblock URL \url{Https://doi.org/10.1371/journal.pone.0178663}.

\bibitem[Cove et~al.(2018)Cove, George, Frascaroli, Brivio, Mayr, Mostafa,
  Indiveri, and Spiga]{Covietal18}
E.~Cove, R.~George, J.~Frascaroli, S.~Brivio, C.~Mayr, H.~Mostafa, G.~Indiveri,
  and S.~Spiga.
\newblock Spike-driven threshold-based learning with memristive synapses and
  neuromorphic silicon neurons.
\newblock \emph{Journal of Physics D: Applied Physics}, 51:\penalty0 344003,
  2018.

\bibitem[Dale et~al.(2016)Dale, Miller, and Trefzer]{Daleetal16}
M.~Dale, S.~Miller, J. F. anbd~Stepney, and M.~A. Trefzer.
\newblock Evolving carbon nanotube reservoir computers.
\newblock In \emph{Proc. Int. Conf. on Unconventional Computation and Natural
  Computation}, pages 49--61, 2016.

\bibitem[Dambre et~al.(2012)Dambre, Verstraeten, Schrauwen, and
  Massar]{Dambreetal12}
J.~Dambre, D.~Verstraeten, B.~Schrauwen, and S.~Massar.
\newblock Information processing capacity of dynamical systems.
\newblock \emph{Nature Scientific Reports}, 2:\penalty0 id 514, 2012.

\bibitem[Davis(1979)]{davis1979behavioural}
William~J Davis.
\newblock Behavioural hierarchies.
\newblock \emph{Trends in Neurosciences}, 2:\penalty0 5--7, 1979.

\bibitem[Doyle et~al.(1998)Doyle, Cabral, Pfuetzner, Kuo, Gulbis, Cohen, Chait,
  and MacKinnon]{doyle1998structure}
Declan~A Doyle, Joao~Morais Cabral, Richard~A Pfuetzner, Anling Kuo,
  Jacqueline~M Gulbis, Steven~L Cohen, Brian~T Chait, and Roderick MacKinnon.
\newblock The structure of the potassium channel: molecular basis of k+
  conduction and selectivity.
\newblock \emph{science}, 280\penalty0 (5360):\penalty0 69--77, 1998.

\bibitem[Durstewitz et~al.(2000)Durstewitz, Seamans, and
  Sejnowski]{Durstewitzetal00}
D.~Durstewitz, J.~K. Seamans, and T.~J. Sejnowski.
\newblock Neurocomputational models of working memory.
\newblock \emph{Nature Neuroscience}, 3:\penalty0 1184--91, 2000.

\bibitem[Eck(2002{\natexlab{a}})]{Eck02}
D.~Eck.
\newblock Real-time musical beat induction with spiking neurons.
\newblock Technical Report IDSIA-22-02, IDSIA, Instituto Dalle Molle di studi
  sull' intelligenza artificiale, Galleria 2, CH-6900 Manno, Switzerland,
  2002{\natexlab{a}}.
\newblock ftp://ftp.idsia.ch/pub/techrep/IDSIA-22-02.ps.gz.

\bibitem[Eck(2002{\natexlab{b}})]{Eck02a}
D.~Eck.
\newblock Finding downbeats with a relaxation oscillator.
\newblock \emph{Psychological Research}, 66\penalty0 (1):\penalty0 18--25,
  2002{\natexlab{b}}.
\newblock
  http://www.iro.umont\-real.ca/$\sim$eckdoug/pa\-pers/2002\_psyres.pdf.

\bibitem[Eck(2007)]{Eck06}
D.~Eck.
\newblock Identifying metrical and temporal structure with an autocorrelation
  phase matrix.
\newblock \emph{Music Perception}, 2007.
\newblock
  http://www.iro.umont\-real.ca/$\sim$eckdoug/pa\-pers/2006\_rppw\_draft.pdf.

\bibitem[Edelman(1993)]{edelman1993neural}
Gerald~M Edelman.
\newblock Neural darwinism: selection and reentrant signaling in higher brain
  function.
\newblock \emph{Neuron}, 10\penalty0 (2):\penalty0 115--125, 1993.

\bibitem[Eliasmith et~al.(2012)Eliasmith, Stewart, Choo, Bekolay, DeWolf, Tang,
  and Rasmussen]{eliasmith2012large}
Chris Eliasmith, Terrence~C Stewart, Xuan Choo, Trevor Bekolay, Travis DeWolf,
  Yichuan Tang, and Daniel Rasmussen.
\newblock A large-scale model of the functioning brain.
\newblock \emph{science}, 338\penalty0 (6111):\penalty0 1202--1205, 2012.

\bibitem[Endrullis et~al.(2019)Endrullis, Klop, and Bakhshi]{Endrullisetal20}
J.~Endrullis, J.~W. Klop, and R.~Bakhshi.
\newblock Transducer degrees: atoms, infima and suprema.
\newblock \emph{Acta Informatica}, 57\penalty0 (3-5):\penalty0 727--758, 2019.

\bibitem[Engels and Sch\"oner(1995)]{EngelsSchoener95}
Ch. Engels and G.~Sch\"oner.
\newblock Dynamic fields endow behavior-based robots with representations.
\newblock \emph{Robotics \& Autonomous Systems}, 14:\penalty0 55--77, 1995.

\bibitem[Ermentrout and Terman(2010)]{Ermentrout2010}
G.B. Ermentrout and D.~Terman.
\newblock \emph{Mathematical Foundations of Neuroscience}, volume~{\bf 35} of
  \emph{Interdisciplinary Applied Mathematics}.
\newblock Springer, New York NY, 2010.

\bibitem[Farhang-Boroujeny(1998)]{FarhangBoroujeny98}
B.~Farhang-Boroujeny.
\newblock \emph{Adaptive Filters: Theory and Applications}.
\newblock Wiley, 1998.

\bibitem[Felleman and Van~Essen(1991)]{felleman1991distributed}
Daniel~J Felleman and David~C Van~Essen.
\newblock Distributed hierarchical processing in the primate cerebral cortex.
\newblock In \emph{Cereb cortex}. Citeseer, 1991.

\bibitem[Fenichel(1979)]{Fenichel1979}
N.~Fenichel.
\newblock Geometric singular perturbation theory for ordinary differential
  equations.
\newblock \emph{Journal of Differential Equations}, {\bf 31}:\penalty0 53--98,
  1979.

\bibitem[Fernando et~al.(2009)Fernando, Liekens, Bingle, Beck, Lenser, Stekel,
  and Rowe]{fernando2009molecular}
Chrisantha~T Fernando, Anthony~ML Liekens, Lewis~EH Bingle, Christian Beck,
  Thorsten Lenser, Dov~J Stekel, and Jonathan~E Rowe.
\newblock Molecular circuits for associative learning in single-celled
  organisms.
\newblock \emph{Journal of the Royal Society Interface}, 6\penalty0
  (34):\penalty0 463--469, 2009.

\bibitem[Forbus(1988)]{Forbus88}
K.~D. Forbus.
\newblock Qualitative physics: past, present and future.
\newblock In \emph{Exploring Artificial Intelligence: Survey Talks from the
  National Conferences on Artificial Intelligence}, pages 239--296. Morgan
  Kaufmann, 1988.

\bibitem[Frady et~al.(2018)Frady, Kleyko, and Sommer]{FradyKleykoSommer18}
E.P. Frady, D.~Kleyko, and F.T. Sommer.
\newblock A theory of sequence indexing and working memory in recurrent neural
  networks.
\newblock \emph{Neural Computation}, 30\penalty0 (6):\penalty0 1449--1513,
  2018.

\bibitem[Franzius et~al.(2008)Franzius, Wilbert, and Wiskott]{Wiskottetal08}
M.~Franzius, B.~Wilbert, and L.~Wiskott.
\newblock Invariant object recognition with slow feature analysis.
\newblock In \emph{Proc. of ICANN 2008}, number 5163 in Lecture Notes in
  Computer Science, pages 961--970. Springer Verlag Berlin, 2008.
\newblock DOI: 10.1007/978-3-540-87536-9\_98.

\bibitem[Fr{\'e}maux and Gerstner(2016)]{fremaux2016neuromodulated}
Nicolas Fr{\'e}maux and Wulfram Gerstner.
\newblock Neuromodulated spike-timing-dependent plasticity, and theory of
  three-factor learning rules.
\newblock \emph{Frontiers in neural circuits}, 9:\penalty0 85, 2016.

\bibitem[French(2003)]{French03}
R.~M. French.
\newblock Catastrophic interference in connectionist networks.
\newblock In L.~Nadel, editor, \emph{Encyclopedia of Cognitive Science},
  volume~1, pages 431--435. Nature Publishing Group, 2003.

\bibitem[Fusi and Wang(2016)]{FusiWang16}
S.~Fusi and X.-J. Wang.
\newblock Short-term, long-term, and working memory.
\newblock In M.~Arbib and J.~Bonaiuto, editors, \emph{From Neuron to Cognition
  via Computational Neuroscience}, pages 319--344. MIT Press, 2016.

\bibitem[Fusi et~al.(2000)Fusi, Annunziato, Badoni, Salamon, and
  Amit]{Fusietal00}
S.~Fusi, M.~Annunziato, D.~Badoni, A.~Salamon, and D.~J. Amit.
\newblock Spike-driven synaptic plasticity: Theory, simulation, {VLSI}
  implementation.
\newblock \emph{Neural Computation}, 12\penalty0 (10):\penalty0 2227--2258,
  2000.

\bibitem[Galtier and Wainrib(2012)]{Galtier2012}
M.~Galtier and G.~Wainrib.
\newblock Multiscale analysis of slow-fast neuronal learning models with noise.
\newblock \emph{Journal of Mathematical Neuroscience}, {\bf 2}\penalty0 (13),
  2012.
\newblock \mbox{doi}: \url{10.1186/2190-8567-2-13}.

\bibitem[Galtier and Wainrib(2013{\natexlab{a}})]{GaltierWainrib13}
M.~Galtier and G.~Wainrib.
\newblock A biological gradient descent for prediction through a combination of
  {STDP} and homeostatic plasticity.
\newblock \emph{Neural Computation}, 25\penalty0 (11):\penalty0 2815--2832,
  2013{\natexlab{a}}.
\newblock URL \url{http://de.arxiv.org/abs/1206.4812}.

\bibitem[Galtier and Wainrib(2013{\natexlab{b}})]{Galtier2013}
M.N. Galtier and G.~Wainrib.
\newblock A biological gradient descent for prediction through a combination of
  {STDP} and homeostatic plasticity.
\newblock \emph{Neural Computation}, {\bf 25}:\penalty0 2815--2832,
  2013{\natexlab{b}}.

\bibitem[Ganguli et~al.(2008)Ganguli, Huh, and
  Sompolinsky]{GanguliHuhSompolinsky08}
S.~Ganguli, D.~Huh, and H.~Sompolinsky.
\newblock Memory traces in dynamical systems.
\newblock \emph{PNAS}, 105\penalty0 (48):\penalty0 18970--18975, 2008.

\bibitem[Gao et~al.(2018)Gao, Zhou, Wang, Cheng, Yachi, and
  Wang]{gao2018dendritic}
Shangce Gao, MengChu Zhou, Yirui Wang, Jiujun Cheng, Hanaki Yachi, and Jiahai
  Wang.
\newblock Dendritic neuron model with effective learning algorithms for
  classification, approximation, and prediction.
\newblock \emph{IEEE transactions on neural networks and learning systems},
  30\penalty0 (2):\penalty0 601--614, 2018.

\bibitem[Garson(2014)]{Garson14}
James Garson.
\newblock Modal logic.
\newblock In Edward~N. Zalta, editor, \emph{The Stanford Encyclopedia of
  Philosophy}. Summer 2014 edition, 2014.

\bibitem[Gerstner et~al.(2018)Gerstner, Lehmann, Liakoni, Corneil, and
  Brea]{gerstner2018eligibility}
Wulfram Gerstner, Marco Lehmann, Vasiliki Liakoni, Dane Corneil, and Johanni
  Brea.
\newblock Eligibility traces and plasticity on behavioral time scales:
  experimental support of neohebbian three-factor learning rules.
\newblock \emph{Frontiers in neural circuits}, 12:\penalty0 53, 2018.

\bibitem[Geuvers et~al.(2010)Geuvers, Koprowski, Synek, and van~der
  Weegen]{Geuversetal10}
H.~Geuvers, A.~Koprowski, D.~Synek, and E.~van~der Weegen.
\newblock Automated machine-checked hybrid system safety proofs.
\newblock In \emph{Proc. Int. Conf. on Interactive Theorem Proving}, pages
  259--274. Springer, 2010.

\bibitem[Graupner and Brunel(2012)]{GraupnerBrunel12}
M.~Graupner and N.~Brunel.
\newblock Calcium-based plasticity model explains sensitivity of synaptic
  changes to spike pattern, rate, and dendritic location.
\newblock \emph{PNAS}, 109\penalty0 (10):\penalty0 3991--3996, 2012.

\bibitem[Gray et~al.(2005)Gray, Hill, and Bargmann]{gray2005circuit}
Jesse~M Gray, Joseph~J Hill, and Cornelia~I Bargmann.
\newblock A circuit for navigation in caenorhabditis elegans.
\newblock \emph{Proceedings of the National Academy of Sciences}, 102\penalty0
  (9):\penalty0 3184--3191, 2005.

\bibitem[Grigoryeva et~al.(2016)Grigoryeva, Henriques, Larger, and
  Ortega]{Grigoryevaetal16}
L.~Grigoryeva, J.~Henriques, L.~Larger, and J.P. Ortega.
\newblock Nonlinear memory capacity of parallel time-delay reservoir computers
  in the processing of multidimensional signals.
\newblock \emph{Neural Computation}, 28\penalty0 (7):\penalty0 1411--1451,
  2016.

\bibitem[Haken(1983)]{Haken83}
H.~Haken.
\newblock \emph{Advanced Synergetics - Instability Hierarchies of
  Self-Organizing Systems and Devices}, volume~20 of \emph{Springer Series in
  Synergetics}.
\newblock Springer, Berlin/Heidelberg, 1983.

\bibitem[Hart et~al.(2019)Hart, Larger, Murphy, and Roy]{Hart2019}
J.D. Hart, L.~Larger, T.E. Murphy, and R.~Roy.
\newblock Delayed dynamical systems: networks, chimeras and reservoir
  computing.
\newblock \emph{Philosophical Transactions of the Royal Society A}, {\bf
  377}:\penalty0 20180123, 2019.

\bibitem[He et~al.(2019{\natexlab{a}})He, J., Galashov, Rusu, Teh, and
  Pascanu]{Heetal19a}
X.~He, Sygnowski J., A.~Galashov, A.~A. Rusu, Y.~W. Teh, and R.~Pascanu.
\newblock Task agnostic continual learning via meta learning.
\newblock In \emph{Proc. neurIPS 2019}, 2019{\natexlab{a}}.
\newblock \url{https://arxiv.org/abs/1906.05201}.

\bibitem[He et~al.(2019{\natexlab{b}})He, Liu, Hadaeghi, and Jaeger]{Heetal19}
X.~He, T.~Liu, F.~Hadaeghi, and H.~Jaeger.
\newblock Reservoir transfer on analog neuromorphic hardware.
\newblock In \emph{Proc. 9th Int. IEEE/EMBS Conf. on Neural Engineering}, pages
  1234--1238, 2019{\natexlab{b}}.

\bibitem[Hebb(1949)]{hebb1949organization}
Donald~Olding Hebb.
\newblock \emph{The organization of behavior: a neuropsychological theory}.
\newblock J. Wiley; Chapman \& Hall, 1949.

\bibitem[Hermans and Schrauwen(2010{\natexlab{a}})]{HermansSchrauwen10}
M.~Hermans and B.~Schrauwen.
\newblock Memory in linear recurrent neural networks in continuous time.
\newblock \emph{Neural Networks}, 23\penalty0 (3):\penalty0 341--355,
  2010{\natexlab{a}}.

\bibitem[Hermans and Schrauwen(2010{\natexlab{b}})]{HermansSchrauwen10a}
M.~Hermans and B.~Schrauwen.
\newblock Memory in reservoirs for high dimensional input.
\newblock In \emph{Proc. WCCI 2010 (IEEE World Congress on Computational
  Intelligence)}, pages 2662--2668, 2010{\natexlab{b}}.

\bibitem[Hermans and Schrauwen(2012)]{HermansSchrauwen11}
M.~Hermans and B.~Schrauwen.
\newblock Recurrent kernel machines: Computing with infinite echo state
  networks.
\newblock \emph{Neural Computation}, 24\penalty0 (1):\penalty0 104--133, 2012.

\bibitem[Hochreiter and Schmidhuber(1997)]{HochreiterSchmidhuber97}
S.~Hochreiter and J.~Schmidhuber.
\newblock Long short-term memory.
\newblock \emph{Neural Computation}, 9\penalty0 (8):\penalty0 1735--1780, 1997.

\bibitem[Hodgkin and Huxley(1952)]{hodgkin1952quantitative}
Alan~L Hodgkin and Andrew~F Huxley.
\newblock A quantitative description of membrane current and its application to
  conduction and excitation in nerve.
\newblock \emph{The Journal of physiology}, 117\penalty0 (4):\penalty0 500,
  1952.

\bibitem[Hopfield(1995)]{hopfield1995pattern}
John~J Hopfield.
\newblock Pattern recognition computation using action potential timing for
  stimulus representation.
\newblock \emph{Nature}, 376\penalty0 (6535):\penalty0 33--36, 1995.

\bibitem[Hopfield and Brody(2001)]{hopfield2001moment}
John~J Hopfield and Carlos~D Brody.
\newblock What is a moment? transient synchrony as a collective mechanism for
  spatiotemporal integration.
\newblock \emph{Proceedings of the National Academy of Sciences}, 98\penalty0
  (3):\penalty0 1282--1287, 2001.

\bibitem[Hyv{\"a}rinen and Oja(1998)]{hyvarinen1998independent}
Aapo Hyv{\"a}rinen and Erkki Oja.
\newblock Independent component analysis by general nonlinear hebbian-like
  learning rules.
\newblock \emph{signal processing}, 64\penalty0 (3):\penalty0 301--313, 1998.

\bibitem[Ibarz et~al.(2011)Ibarz, Casado, and Sanjuan]{Ibarz2011}
B.~Ibarz, J.M. Casado, and M.A.F. Sanjuan.
\newblock Map-based models in neuronal dynamics.
\newblock \emph{Physics Reports}, {\bf 501}:\penalty0 1--74, 2011.

\bibitem[Ivanov and Sharkovsky(1992)]{IvanovSharkovsky1992}
A.F. Ivanov and A.N. Sharkovsky.
\newblock Oscillations in singularly perturbed delay equations.
\newblock In C.K.R.T. Jones, U.~Kirchgraber, and H.O. Walther, editors,
  \emph{Dynamics Reported: Expositions in Dynamical Systems}, volume New
  series: {\bf 1}, pages 164--224. Springer-Verlag, 1992.

\bibitem[Izhikevich(2007)]{Izhikevich2007}
E.M. Izhikevich.
\newblock \emph{Dynamical Systems in Neuroscience: The Geometry of Excitability
  and Bursting}.
\newblock MIT Press, Cambridge MA, 2007.

\bibitem[Izhikevich(2004)]{izhikevich2004model}
Eugene~M Izhikevich.
\newblock Which model to use for cortical spiking neurons?
\newblock \emph{IEEE transactions on neural networks}, 15\penalty0
  (5):\penalty0 1063--1070, 2004.

\bibitem[Jaeger(2002)]{Jaeger02b}
H.~Jaeger.
\newblock Short term memory in echo state networks.
\newblock GMD-Report 152, GMD - German National Research Institute for Computer
  Science, 2002.
\newblock URL
  \url{https://www.ai.rug.nl/minds/uploads/STMEchoStatesTechRep.pdf}.

\bibitem[Jaeger(2014)]{Jaeger14}
H.~Jaeger.
\newblock Controlling recurrent neural networks by conceptors.
\newblock Technical Report~31, Jacobs University Bremen, 2014.
\newblock \href{http://arxiv.org/abs/1403.3369}{arXiv:1403.3369}.

\bibitem[Jaeger(2019)]{Jaeger19a}
H.~Jaeger.
\newblock Computability and complexity.
\newblock Lecture notes of theoretical computer science ii, Jacobs University
  Bremen, 2019.
\newblock \url{https://www.ai.rug.nl/minds/uploads/scriptCC.pdf}.

\bibitem[Jaeger(2020)]{Jaeger20}
H.~Jaeger.
\newblock Exploring the landscapes of “computing”: digital, neuromorphic,
  unconventional --- and beyond.
\newblock manuscript arxiv:2011.12013, 2020.
\newblock URL \url{http://arxiv.org/abs/2011.12013}.

\bibitem[Jaeger and Christaller(1998)]{JaegerChristaller97}
H.~Jaeger and Th. Christaller.
\newblock Dual dynamics: Designing behavior systems for autonomous robots.
\newblock \emph{Artificial Life and Robotics}, 2:\penalty0 108--112, 1998.

\bibitem[Jin and Costa(2015)]{jin2015shaping}
Xin Jin and Rui~M Costa.
\newblock Shaping action sequences in basal ganglia circuits.
\newblock \emph{Current opinion in neurobiology}, 33:\penalty0 188--196, 2015.

\bibitem[Joshi and Triesch(2009)]{JoshiTriesch09}
P.~Joshi and J.~Triesch.
\newblock Optimizing generic neural microcircuits through reward modulated
  {STDP}.
\newblock In \emph{Proc. ICANN 2009}, pages 239--248, 2009.

\bibitem[Jutten and Herault(1991)]{jutten1991blind}
Christian Jutten and Jeanny Herault.
\newblock Blind separation of sources, part i: An adaptive algorithm based on
  neuromimetic architecture.
\newblock \emph{Signal processing}, 24\penalty0 (1):\penalty0 1--10, 1991.

\bibitem[Kairouz et~al.(2019)Kairouz, McMahan, Avent, Bellet, Bennis, Bhagoji,
  Bonawitz, Charles, Cormode, Cummings, D'Oliveira, Rouayheb, Evans, Gardner,
  Garrett, Gasc\'{o}n, Ghazi, Gibbons, Gruteser, Harchaoui, He, He, Huo,
  Hutchinson, Hsu, Jaggi, Javidi, Joshi, Khodak, Kone\v{c}n{y}, Korolova,
  Koushanfar, Koyejo, Lepoint, Liu, Mittal, Mohri, Nock, \"{O}zg\"{u}r, Pagh,
  Raykova, Qi, Ramage, Raskar, Song, Song, Stich, Sun, Suresh, Tram{e}r,
  Vepakomma, Wang, Xiong, Xu, Yang, Yu, Yu, and Zhao]{Kairouzetal19}
Peter Kairouz, H.~Brendan McMahan, Brendan Avent, Aur\'{e}lien Bellet, Mehdi
  Bennis, Arjun~Nitin Bhagoji, K.~A. Bonawitz, Zachary Charles, Graham Cormode,
  Rachel Cummings, Rafael~G.L. D'Oliveira, Salim~El Rouayheb, David Evans, Josh
  Gardner, Zachary Garrett, Adri{a} Gasc\'{o}n, Badih Ghazi, Phillip~B.
  Gibbons, Marco Gruteser, Zaid Harchaoui, Chaoyang He, Lie He, Zhouyuan Huo,
  Ben Hutchinson, Justin Hsu, Martin Jaggi, Tara Javidi, Gauri Joshi, Mikhail
  Khodak, Jakub Kone\v{c}n{y}, Aleksandra Korolova, Farinaz Koushanfar, Sanmi
  Koyejo, Tancr{e}de Lepoint, Yang Liu, Prateek Mittal, Mehryar Mohri, Richard
  Nock, Ayfer \"{O}zg\"{u}r, Rasmus Pagh, Mariana Raykova, Hang Qi, Daniel
  Ramage, Ramesh Raskar, Dawn Song, Weikang Song, Sebastian~U. Stich, Ziteng
  Sun, Ananda~Theertha Suresh, Florian Tram{e}r, Praneeth Vepakomma, Jianyu
  Wang, Li~Xiong, Zheng Xu, Qiang Yang, Felix~X. Yu, Han Yu, and Sen Zhao.
\newblock Advances and open problems in federated learning.
\newblock arxiv report, 2019.
\newblock URL \url{https://arxiv.org/abs/1912.04977}.

\bibitem[Kandel et~al.(2014)Kandel, Dudai, and Mayford]{kandel2014molecular}
Eric~R Kandel, Yadin Dudai, and Mark~R Mayford.
\newblock The molecular and systems biology of memory.
\newblock \emph{Cell}, 157\penalty0 (1):\penalty0 163--186, 2014.

\bibitem[Kaper(1999)]{Kaper1999}
T.J. Kaper.
\newblock An introduction to geometric methods and dynamical systems theory for
  singular perturbation problems.
\newblock \emph{Proceedings of Symposia in Applied Mathematics}, {\bf
  56}:\penalty0 85--131, 1999.

\bibitem[Kaplan et~al.(2020)Kaplan, Thula, Khoss, and Zimmer]{kaplan2020nested}
Harris~S Kaplan, Oriana~Salazar Thula, Niklas Khoss, and Manuel Zimmer.
\newblock Nested neuronal dynamics orchestrate a behavioral hierarchy across
  timescales.
\newblock \emph{Neuron}, 105\penalty0 (3):\penalty0 562--576, 2020.

\bibitem[Katz(1971)]{katz1971quantal}
B~Katz.
\newblock Quantal mechanism of neural transmitter release.
\newblock \emph{Science}, 173\penalty0 (3992):\penalty0 123--126, 1971.

\bibitem[Klampfl and Maass(2013)]{KlampflMaass13}
S.~Klampfl and W.~Maass.
\newblock Emergence of dynamic memory traces in cortical microcircuit models
  through {STDP}.
\newblock \emph{J. of Neuroscience}, 33\penalty0 (28):\penalty0 11515, 2013.

\bibitem[Kohonen(1982)]{kohonen1982self}
Teuvo Kohonen.
\newblock Self-organized formation of topologically correct feature maps.
\newblock \emph{Biological cybernetics}, 43\penalty0 (1):\penalty0 59--69,
  1982.

\bibitem[Kokotovic et~al.(1999)Kokotovic, Khalil, and O'Reilly]{Kokotovic1999}
P.~Kokotovic, H.K. Khalil, and J.~O'Reilly.
\newblock \emph{Singular Perturbation Methods in Control Analysis and Design},
  volume~{\bf 25} of \emph{Classics in Applied Mathematics}.
\newblock SIAM, Philadelphia PA, 1999.

\bibitem[Kuehn(2015)]{Kuehn2015}
C.~Kuehn.
\newblock \emph{Multiple Time Scale Dynamics}.
\newblock Springer, Cham, Switzerland, 1st edition, 2015.

\bibitem[Kusmierz et~al.(2017)Kusmierz, Isomura, and Toyoizumi]{Kusmierzetal17}
L.~Kusmierz, T.~Isomura, and T.~Toyoizumi.
\newblock Learning with three factors: modulating hebbian plasticity with
  errors.
\newblock \emph{Current Opinion in Neurobiology}, 46:\penalty0 170--177, 2017.

\bibitem[Lane et~al.(2015)Lane, Kanjlia, Omaki, and Bedny]{lane2015visual}
Connor Lane, Shipra Kanjlia, Akira Omaki, and Marina Bedny.
\newblock “visual” cortex of congenitally blind adults responds to
  syntactic movement.
\newblock \emph{Journal of Neuroscience}, 35\penalty0 (37):\penalty0
  12859--12868, 2015.

\bibitem[Large and Palmer(2002)]{LargePalmer02}
E.~W. Large and C.~Palmer.
\newblock Perceiving temporal regularity in music.
\newblock \emph{Cognitive Science}, 26:\penalty0 1--37, 2002.

\bibitem[Legenstein and Maass(2007)]{LegensteinMaass07}
R.~Legenstein and W.~Maass.
\newblock Edge of chaos and prediction of computational performance for neural
  circuit models.
\newblock \emph{Neural Networks}, 20\penalty0 (3):\penalty0 323--334, 2007.

\bibitem[Linaro et~al.(2012)Linaro, Champneys, Desroches, and
  Storace]{Linaro2012}
D.~Linaro, A.~Champneys, M.~Desroches, and M.~Storace.
\newblock Codimension-two homoclinic bifurcations underlying spike adding in
  the {Hindmarsh-Rose} burster.
\newblock \emph{SIAM Journal on Applied Dynamical Systems}, {\bf 11}\penalty0
  (3):\penalty0 939--962, 2012.

\bibitem[Lins and Sch\"{o}ner(2014)]{LinsSchoener14}
J.~Lins and G.~Sch\"{o}ner.
\newblock Neural fields.
\newblock In S.~Coombes, P.~beim Graben, R.~Potthast, and J.~Wright, editors,
  \emph{Neural fields: theory and applications}, pages 319--339. Springer
  Verlag, 2014.

\bibitem[Long et~al.(2010)Long, Jin, and Fee]{long2010support}
Michael~A Long, Dezhe~Z Jin, and Michale~S Fee.
\newblock Support for a synaptic chain model of neuronal sequence generation.
\newblock \emph{Nature}, 468\penalty0 (7322):\penalty0 394--399, 2010.

\bibitem[Luko\v{s}evi\v{c}ius et~al.(2006)Luko\v{s}evi\v{c}ius, Popovici,
  Jaeger, and Siewert]{Lukoseviciusetal06}
M.~Luko\v{s}evi\v{c}ius, D.~Popovici, H.~Jaeger, and U.~Siewert.
\newblock Time warping invariant echo state networks.
\newblock IUB Technical Report~2, International University Bremen, 2006.
\newblock URL \url{https://www.ai.rug.nl/minds/uploads/techreport2.pdf}.

\bibitem[Maass et~al.(2002)Maass, Natschl{\"a}ger, and Markram]{Maassetal01}
W.~Maass, T.~Natschl{\"a}ger, and H.~Markram.
\newblock Real-time computing without stable states: A new framework for neural
  computation based on perturbations.
\newblock \emph{Neural Computation}, 14\penalty0 (11):\penalty0 2531--2560,
  2002.
\newblock http://www.cis.tugraz.at/igi/maass/psfiles/LSM-v106.pdf.

\bibitem[Markram et~al.(1997)Markram, L\"{u}bke, Frotscher, and
  Sakmann]{Markrametal97}
H.~Markram, J.~L\"{u}bke, M.~Frotscher, and B.~Sakmann.
\newblock Regulation of synaptic efficacy by coincidence of postsynaptic {APs}
  and {EPSPs}.
\newblock \emph{Science}, 275\penalty0 (5297):\penalty0 213--215, 1997.

\bibitem[Minnai et~al.(2018)Minnai, Mirigliano, Brown, and
  Milani]{Minnaietal18}
C.~Minnai, M.~Mirigliano, S.~A. Brown, and P.~Milani.
\newblock The nanocoherer: an electrically and mechanically resettable
  resistive switching device based on gold clusters assembled on paper.
\newblock \emph{Nano Futures}, 2:\penalty0 011002, 2018.

\bibitem[Mira and Shilnikov(2005)]{MiraShilnikov2005}
C.~Mira and A.~Shilnikov.
\newblock Slow-fast dynamics generated by noninvertible plane maps.
\newblock \emph{International Journal of Bifurcation and Chaos}, {\bf
  115}\penalty0 (11):\penalty0 3509--3534, 2005.

\bibitem[Moradi et~al.(2018)Moradi, Qiao, Stefanini, and
  Indiveri]{Moradietal18}
S.~Moradi, N.~Qiao, F.~Stefanini, and G.~Indiveri.
\newblock A scalable multicore architecture with heterogeneous memory
  structures for dynamic neuromorphic asynchronous processors (dynaps).
\newblock \emph{IEEE Transactions on Biomedical Circuits and Systems},
  12\penalty0 (1):\penalty0 106--122, 2018.

\bibitem[Neal(1993)]{Neal93}
R.M. Neal.
\newblock Probabilistic inference using {Markov} chain {Monte Carlo} methods.
\newblock Technical Report CRG-TR-93-1, Dpt. of Computer Science, University of
  Toronto, 1993.

\bibitem[Nefti and Averbeck(2019)]{NeftciAverbeck19}
E.~O. Nefti and B.~B. Averbeck.
\newblock Reinforcement learning in artificial and biological systems.
\newblock \emph{Nature Machine Intelligence}, 1\penalty0 (3):\penalty0
  133--143, 2019.

\bibitem[Oh et~al.(2015)Oh, Parajuli, and Zito]{oh2015heterosynaptic}
Won~Chan Oh, Laxmi~Kumar Parajuli, and Karen Zito.
\newblock Heterosynaptic structural plasticity on local dendritic segments of
  hippocampal ca1 neurons.
\newblock \emph{Cell reports}, 10\penalty0 (2):\penalty0 162--169, 2015.

\bibitem[Oja(1982)]{oja1982simplified}
Erkki Oja.
\newblock Simplified neuron model as a principal component analyzer.
\newblock \emph{Journal of mathematical biology}, 15\penalty0 (3):\penalty0
  267--273, 1982.

\bibitem[{O'Malley, Jr.}(1991)]{OMalley1991}
R.E. {O'Malley, Jr.}
\newblock \emph{Singular Perturbation Methods for Ordinary Differential
  Equations}, volume~{\bf 89} of \emph{Applied Mathematical Sciences}.
\newblock Springer-Verlag, New York NY, 1991.

\bibitem[Parisi et~al.(2019)Parisi, Kemker, Part, Kanan, and
  Wermter]{Parisietal19}
G.~I. Parisi, R~Kemker, J.~L. Part, C.~Kanan, and S.~Wermter.
\newblock Continual lifelong learning with neural networks: A review.
\newblock \emph{Neural Networks}, 113:\penalty0 54--71, 2019.

\bibitem[Pascanu and Jaeger(2011)]{PascanuJaeger10}
R.~Pascanu and H.~Jaeger.
\newblock A neurodynamical model for working memory.
\newblock \emph{Neural Networks}, 24\penalty0 (2):\penalty0 199--207, 2011.
\newblock DOI: 10.1016/j.neunet.2010.10.003.

\bibitem[Payvand et~al.(2020)Payvand, Kurdahi, Eltawil, and
  Neftci]{Payvandetal20}
M.E. Payvand, M. abnd~Fouda, F.~Kurdahi, A.~Eltawil, and E.O. Neftci.
\newblock Error-triggered three-factor learning dynamics for crossbar arrays.
\newblock In \emph{Proc. 2nd IEEE International Conference on Artificial
  Intelligence Circuits and Systems (AICAS)}, pages 218--222, 2020.

\bibitem[Piaget(1952)]{piaget1952origins}
Jean Piaget.
\newblock \emph{The origins of intelligence in children}.
\newblock International Universities Press New York, 1952.

\bibitem[Prychynenko et~al.(2018)Prychynenko, Sitte, Litzius, Kr\"{u}ger,
  Bourianoff, Kl\"{a}ui, Sinova, and Everschor-Sitte]{Prychynenkoetal18}
D.~Prychynenko, M.~Sitte, K.~Litzius, B.~Kr\"{u}ger, G.~Bourianoff,
  M.~Kl\"{a}ui, J.~Sinova, and K.~Everschor-Sitte.
\newblock Magnetic skyrmion as a nonlinear resistive element: A potential
  building block for reservoir computing.
\newblock \emph{Phys. Rev. Applied}, 9:\penalty0 014034, 2018.

\bibitem[Pusuluri et~al.(2020)Pusuluri, Ju, and Shilnikov]{Pusuluri2020}
K.~Pusuluri, H.~Ju, and A.L. Shilnikov.
\newblock Chaotic dynamics in neural systems.
\newblock In R.A. Meyers, editor, \emph{Encyclopedia of Complexity and Systems
  Science}, page : to appear. Springer Science, 2020.

\bibitem[Rall(2009)]{rall2009rall}
Wilfrid Rall.
\newblock Rall model.
\newblock \emph{Scholarpedia}, 4\penalty0 (4):\penalty0 1369, 2009.

\bibitem[Ranjan et~al.(2011)Ranjan, Khazen, Gambazzi, Ramaswamy, Hill,
  Sch{\"u}rmann, and Markram]{ranjan2011channelpedia}
Rajnish Ranjan, Georges Khazen, Luca Gambazzi, Srikanth Ramaswamy, Sean~L Hill,
  Felix Sch{\"u}rmann, and Henry Markram.
\newblock Channelpedia: an integrative and interactive database for ion
  channels.
\newblock \emph{Frontiers in neuroinformatics}, 5:\penalty0 36, 2011.

\bibitem[Roclin et~al.(2013)Roclin, Bichler, Gamrat, Thorpe, and
  Klein]{Roclinetal13}
D.~Roclin, O.~Bichler, C.~Gamrat, S.~J. Thorpe, and J.~O. Klein.
\newblock Design study of efficient digital order-based {STDP} neuron
  implementations for extracting temporal features.
\newblock In \emph{Proc. of The 2013 International Joint Conference on Neural
  Networks (IJCNN)}, pages 1--7, 2013.

\bibitem[Rubin and Terman(2002)]{RubinTerman2002}
J.E. Rubin and D.~Terman.
\newblock Chapter 3: Geometric singular perturbation analysis of neuronal
  dynamics.
\newblock In B.~Fiedler, editor, \emph{Handbook of Dynamical Systems},
  volume~{\bf 2}, pages 93--146. Elsevier, 2002.

\bibitem[Ruschel(2020)]{Ruschel2020}
S.~Ruschel.
\newblock \emph{Multiple Time-Scale Delay Systems in Mathematical Biology and
  Laser Dynamics}.
\newblock PhD thesis, Technical University of Berlin, 2020.

\bibitem[Saffiotti et~al.(1995)Saffiotti, Konolige, and
  Ruspini]{Saffiottietal95}
A.~Saffiotti, K.~Konolige, and E.H. Ruspini.
\newblock A multivalued logic approach to integrating planning and control.
\newblock \emph{Artificial Intelligence}, 76:\penalty0 481--526, 1995.

\bibitem[Sch\"{o}nfeld and Wiskott(2015)]{SchoenfeldWiskott15}
F.~Sch\"{o}nfeld and L.~Wiskott.
\newblock Modeling place field activity with hierarchical slow feature
  analysis.
\newblock \emph{Frontiers in Computational Neuroscience}, 9:\penalty0 51, 2015.

\bibitem[Schrauwen et~al.(2008)Schrauwen, Wardermann, Verstraeten, Steil, and
  Stroobandt]{Schrauwenetal08}
Benjamin Schrauwen, Marion Wardermann, David Verstraeten, Jochen~J. Steil, and
  Dirk Stroobandt.
\newblock Improving reservoirs using intrinsic plasticity.
\newblock \emph{Neurocomputing}, 71:\penalty0 1159--1171, 1 2008.

\bibitem[Sheldon et~al.(2020)Sheldon, Kolchinsky, and Caravelli]{Sheldonetal20}
F.~C. Sheldon, A.~Kolchinsky, and F.~Caravelli.
\newblock The computational capacity of memristor reservoirs.
\newblock Arxiv manuscript, 2020.
\newblock URL \url{https://arxiv.org/abs/2009.00112v2}.

\bibitem[Shouval et~al.(2002)Shouval, Bear, and Cooper]{shouval2002unified}
Harel~Z Shouval, Mark~F Bear, and Leon~N Cooper.
\newblock A unified model of nmda receptor-dependent bidirectional synaptic
  plasticity.
\newblock \emph{Proceedings of the National Academy of Sciences}, 99\penalty0
  (16):\penalty0 10831--10836, 2002.

\bibitem[Siegelmann and Sontag(1994)]{SiegelmannSontag94}
H.T. Siegelmann and E.D. Sontag.
\newblock Analog computation via neural network.
\newblock \emph{Theoretical Computer Science}, 131\penalty0 (2):\penalty0
  331--360, 1994.

\bibitem[Sima and Orponen(2003)]{SimaOrponen03}
J.~Sima and P.~Orponen.
\newblock General-purpose computation with neural networks: A survey of
  complexity theoretic results.
\newblock \emph{Neural Computation}, 15\penalty0 (12):\penalty0 2727--2778,
  2003.

\bibitem[Skarda(1999)]{skarda1999perceptual}
Christine~A Skarda.
\newblock The perceptual form of life.
\newblock \emph{Journal of Consciousness Studies}, 6\penalty0 (11-12):\penalty0
  79--93, 1999.

\bibitem[Song et~al.(2000)Song, Miller, and Abbott]{Songetal00}
S.~Song, K.~D. Miller, and L.~F. Abbott.
\newblock Competitive hebbian learning through spike-timing-dependent synaptic
  plasticity.
\newblock \emph{Nature Neuroscience}, 3\penalty0 (9):\penalty0 919--926, 2000.

\bibitem[Stanley et~al.(2019)Stanley, Clune, Lehman, and
  Miikkulainen]{stanley2019designing}
Kenneth~O Stanley, Jeff Clune, Joel Lehman, and Risto Miikkulainen.
\newblock Designing neural networks through neuroevolution.
\newblock \emph{Nature Machine Intelligence}, 1\penalty0 (1):\penalty0 24--35,
  2019.

\bibitem[Strauss et~al.(2012)Strauss, Wustlich, and Labahn]{Straussetal12}
T.~Strauss, W.~Wustlich, and R.~Labahn.
\newblock Design strategies for weight matrices of echo state networks.
\newblock \emph{Neural Computation}, 24\penalty0 (12):\penalty0 3246--3276,
  2012.

\bibitem[Subramoney et~al.(2021)Subramoney, Scherr, and
  Maass]{Subramoneyetal21}
A.~Subramoney, F.~Scherr, and W.~Maass.
\newblock Reservoirs learn to learn.
\newblock In K.~Nakajima and I.~Fischer, editors, \emph{Reservoir Computing:
  Theory, Physical Implementations and Applications}, chapter 1.3. Springer,
  2021.
\newblock Preprint arXiv:1909.07486.

\bibitem[Sutton and Barto(1998)]{SuttonBarto98}
R.~Sutton and A.~G. Barto.
\newblock \emph{Reinforcement learning: an introduction}.
\newblock Cambridge: MIT press, 1998.
\newblock online version at
  \url{http://incompleteideas.net/book/ebook/the-book.html}.

\bibitem[Tanaka and et~al(2019)]{Tanakaetal18}
G.~Tanaka and et~al.
\newblock Recent advances in physical reservoir computing: A review.
\newblock \emph{Neural Networks}, 115:\penalty0 100--123, 2019.
\newblock preprint in \url{https://arxiv.org/abs/1808.04962}.

\bibitem[Terman(1991)]{Terman1991}
D.~Terman.
\newblock Chaotic spikes arising from a model of bursting in excitable cell
  membranes.
\newblock \emph{SIAM Journal of Applied Mathematics}, {\bf 51}\penalty0
  (5):\penalty0 1418--1450, 1991.

\bibitem[Thiele et~al.(2018)Thiele, Bichler, and Dupret]{Thieleetal18}
J.~C. Thiele, O.~Bichler, and A.~Dupret.
\newblock Event-based, timescale invariant unsupervised online deep learning
  with {STDP}.
\newblock \emph{Frontiers in Computational Neuroscience}, 12:\penalty0 article
  46, 2018.

\bibitem[Tinbergen(1951)]{tinbergen1951study}
Niko Tinbergen.
\newblock \emph{The study of instinct.}
\newblock Clarendon Press/Oxford University Press, 1951.

\bibitem[Torrejon et~al.(2017)Torrejon, Riou, Tsunegi, Khalsa, Querlioz,
  Bortolotti, Cros, Yakushiji, Fukushima, Kubota, Yuasa, Stiles, and
  Grollier]{Torrejonetal17}
J.~Torrejon, F.~A. Riou, M.~Araujo, S.~Tsunegi, G.~Khalsa, D.~Querlioz,
  P.~Bortolotti, V.~Cros, K.~Yakushiji, A.~Fukushima, H.~Kubota, S.~Yuasa,
  M.~D. Stiles, and J.~Grollier.
\newblock Neuromorphic computing with nanoscale spintronic oscillators.
\newblock \emph{Nature}, 547\penalty0 (27 July):\penalty0 428--431, 2017.

\bibitem[Van~Kesteren et~al.(2012)Van~Kesteren, Ruiter, Fern{\'a}ndez, and
  Henson]{van2012schema}
Marlieke~TR Van~Kesteren, Dirk~J Ruiter, Guill{\'e}n Fern{\'a}ndez, and
  Richard~N Henson.
\newblock How schema and novelty augment memory formation.
\newblock \emph{Trends in neurosciences}, 35\penalty0 (4):\penalty0 211--219,
  2012.

\bibitem[van Leeuwen and Wiedermann(2001)]{vanLeeuwenWiedermann01}
J.~van Leeuwen and J.~Wiedermann.
\newblock Beyond the turing limit: Evolving interactive systems.
\newblock In \emph{International Conference on Current Trends in Theory and
  Practice of Computer Science}, number 2234 in LNCS, pages 90--109. Springer,
  2001.

\bibitem[Vasil'eva and Volosov(1967)]{Vasileva1967}
A.B. Vasil'eva and V.M. Volosov.
\newblock The work of {Tikhonov} and his pupils on ordinary differential
  equations containing a small parameter.
\newblock \emph{Russian Mathematical Surveys}, {\bf 22}:\penalty0 124--142,
  1967.

\bibitem[Verhulst(2005)]{Verhulst2005}
F.~Verhulst.
\newblock \emph{Methods and Applications of Singular Perturbations: Boundary
  Layers and Multiple Timescale Dynamics}, volume~{\bf 50} of \emph{Texts in
  Applied Mathematics}.
\newblock Springer, New York NY, 2005.

\bibitem[Von~der Malsburg(1973)]{von1973self}
Chr Von~der Malsburg.
\newblock Self-organization of orientation sensitive cells in the striate
  cortex.
\newblock \emph{Kybernetik}, 14\penalty0 (2):\penalty0 85--100, 1973.

\bibitem[Von~Frisch(1967)]{von1967dance}
Karl Von~Frisch.
\newblock \emph{The dance language and orientation of bees.}
\newblock Harvard University Press, 1967.

\bibitem[Walker and Stickgold(2006)]{walker2006sleep}
Matthew~P Walker and Robert Stickgold.
\newblock Sleep, memory, and plasticity.
\newblock \emph{Annu. Rev. Psychol.}, 57:\penalty0 139--166, 2006.

\bibitem[Wiskott and Sejnowski(2002)]{WiskottSejnowski02}
L.~Wiskott and T.~Sejnowski.
\newblock Slow feature analysis: Unsupervised learning of invariances.
\newblock \emph{Neural Computation}, 14\penalty0 (4):\penalty0 715--770, 2002.

\bibitem[wyffels et~al.(2014)wyffels, Li, Waegeman, Schrauwen, and
  Jaeger]{wyffelsetal13}
F.~wyffels, J.~Li, T.~Waegeman, B.~Schrauwen, and H.~Jaeger.
\newblock Frequency modulation of large oscillatory neural networks.
\newblock \emph{Biological Cybernetics}, 108:\penalty0 145--157, 2014.

\bibitem[Yamashita and Tani(2008)]{YamashitaTani08}
Y.~Yamashita and J.~Tani.
\newblock Emergence of functional hierarchy in a multiple timescale neural
  network model: A humanoid robot experiment.
\newblock \emph{PLOS Computational Biology}, 4\penalty0 (11):\penalty0
  e1000220, 2008.

\bibitem[Yanchuk and Giacomelli(2017)]{Yanchuk2017}
S.~Yanchuk and G.~Giacomelli.
\newblock Spatio-temporal phenomena in complex systems with time delays.
\newblock \emph{Journal of Physics A: Mathematical and Theoretical}, {\bf
  50}:\penalty0 103001, 2017.

\bibitem[Yousefzadeh et~al.(2018)Yousefzadeh, Stromatias, Soto,
  Serrano-Gotarredona, and Linares-Barranco]{Yousefzadehetal18}
A.~Yousefzadeh, E.~Stromatias, M.~Soto, T.~Serrano-Gotarredona, and
  B.~Linares-Barranco.
\newblock On practical issues for stochastic {STDP} hardware with 1-bit
  synaptic weights.
\newblock \emph{Frontiers in Neuroscience}, 12\penalty0 (October):\penalty0
  Article Nr 665, 2018.

\end{thebibliography}
\end{document}